\documentclass[12pt]{article}
\usepackage{amssymb,amsmath,epsfig}
\usepackage{graphicx}
\usepackage{amsfonts,graphicx,amssymb,amsmath,amsthm,epsfig}
\usepackage{float}
\allowdisplaybreaks
\begin{document}
\title{\textbf{The Dynamics of Cosmic Evolution: Insights from Bouncing Cosmology}}
\author{M. Sharif$^{1}$\thanks{msharif.math@pu.edu.pk} ,
M.Zeeshan Gul$^{1,2}$\thanks{mzeeshangul.math@gmail.com}
~and Ahmad Nawaz$^{1}$\thanks{ahmadch.math@gmail.com}\\
$^1$ Department of Mathematics and Statistics, The University of
Lahore,\\ 1-KM Defence Road Lahore-5400, Pakistan.\\
$^2$ Research Center of Astrophysics and Cosmology, Khazar
University,\\ Baku, AZ1096, 41 Mehseti Street, Azerbaijan.}

\date{}
\maketitle

\begin{abstract}
The primary aim of this work is to explore feasible bouncing
cosmological solutions in the framework of $f(\mathcal{Q},
\mathcal{C})$ gravity, where $\mathcal{Q}$ denotes non-metricity and
$\mathcal{C}$ indicates the boundary term. To achieve this, we
analyze the dynamics of a Bianchi type-I spacetime with perfect
fluid distribution. We consider various functional forms of
$f(\mathcal{Q,C})$ theory to assess how this modified gravity
framework influences cosmic evolution. Additionally, we examine the
dynamics of different cosmological parameters to explore
non-singular bounce solutions. We also use linear perturbation to
study the stability analysis. Our findings reveal the breach of the
null energy conditions, which is required for the existence of
viable bounce solutions. The equation of state parameter
demonstrates either a quintessence phase or a phantom regime of the
universe, demonstrating that the cosmos is undergoing accelerating
expansion. This gravitational framework presents a promising
alternative to the standard cosmological model, presenting an
innovative viewpoint on gravitational interactions and the dynamics
of the early universe.
\end{abstract}

\section{Introduction}

Einstein gravitational theory $(\mathcal{EGT})$ is a geometric
framework which serves as a fundamental cornerstone of modern
physics. Various cosmic observations such as supernovae type Ia,
suggest that the universe is now undergoing expansion \cite{1}. The
universe is composed mainly of three elements,i.e., dark energy
($\mathcal{DE}$), dark matter ($\mathcal{DM}$) and ordinary matter.
Dark energy, a non-baryonic, invisible form of energy, is inferred
from observations like gravitational lensing and galaxy rotation
curves \cite{2}. The accelerated expansion of the universe, one of
the most notable findings in recent cosmological research has
directed scientific focus toward unknown realms. This expansion is
thought to be driven by an unidentified energy source with strong
negative pressure, commonly referred to as $\mathcal{DE}$. The
origins and properties of $\mathcal{DE}$ make it one of the most
significant and unsolved enquiries in cosmology. The standard cold
dark matter ($\Lambda\mathcal{CDM}$) model is currently the leading
framework to describe the nature of $\mathcal{DE}$. In this model,
the cosmological constant $(\Lambda)$ is a key candidate for
explaining cosmic acceleration. However, despite its efficacy in
aligning with empirical data, the model has significant challenges,
including the fine-tuning and cosmic coincidence difficulties
\cite{3}-\cite{7}.

To investigate the origin of cosmic accelerated expansion, two
approaches are generally employed. The first approach alters the
geometric aspect of the Einstein-Hilbert action, leading to modified
theories of gravity ($\mathcal{MGT}s$). The second approach focuses
on modifying the matter component, resulting in models of
$\mathcal{DE}$ \cite{8}. Alternative frameworks such as
$\mathcal{MGT}s$ aim to address the challenges related to
$\mathcal{DE}$. We use the term modified to describe the changes or
additions to the geometric aspect of the Einstein-Hilbert action. We
can use equivalent geometric frameworks to depict $\mathcal{EGT}$.
The first is the curvature representation, where non-metricity and
torsion are absent. The second approach is the teleparallel
formalism, which eliminates both non-metricity and curvature. An
alternative formulation interprets gravitational interactions via
the non-metricity, which accounts for variations in the lengths of
vectors during parallel transport. Teleparallel gravity is based on
a collection of tetrad vectors rather than a metric tensor,
resulting in the manifestation of torsion. In this framework,
curvature is defined by torsion produced by the tetrads, providing a
way to explain the gravitational phenomena observed in the universe.
Jimenez et al \cite{10} introduced the idea of symmetric
teleparallel gravity, which is called $f(\mathcal{Q})$ gravity. A
detailed analysis of the geometric and physical aspects of this
gravity has been studied in \cite{10aa}-\cite{10hh} to provide
valuable insights for understanding its principles. There are
different forms of $\mathcal{MGT}s$ such as curvature, torsion and
non-metricity based theories \cite{11}-\cite{12}.

De et al  \cite{13} extended the symmetric teleparalell theory by
including the boundary term in the functional action, known as
$f(\mathcal{Q}, \mathcal{C})$ theory. This modified proposal has
attracted significant interest among researchers because of its
fascinating consequences in gravitational physics \cite{14}. This
alternative framework has been subject of great interest due to its
theoretical ramifications and significance in astrophysical and
cosmological contexts. The incorporation of non-metricity and the
existence of boundary terms in this theory allow for a more
intricate representation of gravitational interactions.
Non-metricity refers to the departure from the Levi-Civita
connection, which is the connection compatible with the metric
tensor $\mathcal{EGT}$. This theory gives the significant impact on
the gravitational field and provides novel gravitational dynamics
and cosmological solutions because of coupling of non-metricity and
boundary term. Thus, the objective of this theory is to provide a
more extensive framework for gravitational physics and cosmology
that may effectively resolves existing empirical inconsistencies and
provides valuable insights into fundamental inquiries about the
characteristics of gravity and the universe.

The structure of $f(\mathcal{Q}, \mathcal{C})$ theory is inspired by
other well known modifications of gravity such as $f(\mathcal{R})$,
$f(T)$ and $f(\mathcal{Q})$ theories. These models have gained
considerable interest due to their ability to extend $\mathcal{EGT}$
while still adhering to many of the experimental and observational
constraints. The inclusion of the boundary term adds flexibility and
allows for a broader range of viable solutions. The combination of
non-metricity and the boundary term allows for the construction of a
geometrically consistent theory that maintains key phenomenological
advantages such as explaining late time acceleration and producing
viable inflationary models. Thus, this $\mathcal{MGT}$ is motivated
by the desire to go beyond the limitations of $\mathcal{EGT}$ to
resolve cosmological singularities and explore novel geometric
structures through non-metricity, providing a unified explanation
for $\mathcal{DE}$, dark matter and cosmic inflation. The inclusion
of the boundary term enriches the theoretical framework, offering
new avenues for cosmological model building and consistency with
current observations \cite{14a}.

The motivation behind considering a modified $f(\mathcal{Q},
\mathcal{C})$ theory arises from several key factors related to both
theoretical considerations and observational needs in cosmology and
gravitational physics. The boundary term plays an important role in
the action formulation of $\mathcal{MGT}$. In $\mathcal{EGT}$, the
boundary term is typically related to the Gibbons-Hawking-York
boundary term, which does not affect the equations of motion but
ensures a well posed variational principle. Boundary term
contributes to the dynamical equations, leading to new insights into
cosmological evolution, effective gravitational constants or
corrections to the standard cosmological model. This alternative
theory offers a framework to explain cosmic acceleration without
invoking a cosmological constant, addressing the $\mathcal{DE}$
problem.

Maurya \cite{14b} analyzed an isotropic and homogeneous flat dark
energy model in $f(\mathcal{Q},\mathcal{C})$ gravity. Maurya
\cite{14c} investigated $\mathcal{DE}$ cosmological models in non
coincident gauge formulation of non-metricity gravity with boundary
term. Usman et al \cite{14d} examined the disparity between matter
and antimatter in the universe with the help of gravitational
baryogenesis in $f(\mathcal{Q},\mathcal{C})$ theory. Sadatian et al
\cite{14e} studied the relationship between cosmological inflation
and slow roll parameters, contributing to our understanding of the
dynamics and implications of inflation in the early universe
according to the modified $f(\mathcal{Q},\mathcal{C})$ model.
Samaddar et al \cite{14f} considered Holographic and Renyi
holographic $\mathcal{DE}$ models to examine the cosmic evolution in
this $\mathcal{MGT}$. This modified framework provides an effective
description of the Universe evolutionary history and fits well with
contemporary cosmic data \cite{14g}. The matter, anti matter
asymmetry in the early Universe history is explained in \cite{14h}.
It is found that this approach allows for the exploration of cosmic
evolution and the derived models are physically viable for the
observed baryon to entropy ratio value.

The widely accepted big bang theory posits that the universe
originated from the initial singularity, accounting for many
observable cosmic structures. However, this model encounters certain
challenges include the horizon and flatness issues \cite{15}.
Inflationary theory was introduced to address these issues,
providing a mechanism for early expansion \cite{16}. Although, this
theory resolves many of these issues, but it does not eliminate the
singularity problem. In this response, alternative models like
bouncing cosmology have been developed to tackle this challenge.
Bouncing cosmology posits that the cosmos experiences cyclic stages
of expansion and contraction, in contrast to the big bang theory,
thereby avoiding the initial singularity \cite{17}. Bouncing
cosmological models are promising, as they address the singularity
issue and resolve the horizon and flatness problems.

When the cosmos exhibits neither homogeneity nor isotropy, the
complexity of modeling its behavior significantly increases. In such
scenarios, the cosmos cannot be described by simple models like FRW.
Instead, we need to consider more complex and less symmetric models
like Bianchi type-I that can account for spatial variations in
density, pressure and other cosmological parameters. The
anisotropies and inhomogeneities would affect the cosmic microwave
background radiation and provide clues about the early cosmic
conditions and the processes that led to its current state. The
dynamical evolution of cosmos including its expansion rate and
potential fate would also influenced by these irregularities. Thus,
exploring alternative models like Bianchi type-I is crucial for
understanding the cosmic anisotropic features and it lead to a more
comprehensive and accurate evolution of cosmos.

Bianchi type-I is the simplest anisotropic generalization of the
FLRW model. It has a straightforward mathematical formulation,
making it a suitable starting point for exploring the effects of
anisotropy. It introduces anisotropy while retaining a flat spatial
geometry. This allows us to isolate and study the influence of
anisotropy without the confounding effects of spatial curvature,
which are present in other Bianchi types. The Bianchi type-I model
provides a rich background for comparison and validation of our
results. By using this well established model, one can build upon
previous work and offer a clear context for our findings.
Observational data such as the cosmic microwave background suggests
that if there is any deviation from isotropy, it is likely to be
small. The Bianchi type-I model captures this small anisotropy
effectively without introducing more complex anisotropic features
that might be harder to detect or constrain observationally.

The primary goal of the Bianchi universe model is to go beyond the
isotropic FRW model in order to examine the dynamics of both
anisotropic and homogeneous cosmos. This method provides significant
perspectives on the formation of the early universe, cosmic
formations and how gravitational fields behave under different
circumstances. The model encompasses spatial anisotropy while
upholding uniformity and yields profound understanding of cosmic
progression. Bianchi universe models deviate from uniformity while
sustaining homogeneity. This departure allows for spatial anisotropy
and suggests that properties of the universe can vary in different
directions. Scientists can use the behavior of anisotropic universe
to enhance our current understanding of gravity. Bianchi models
offer a framework for investigating how inflationary dynamics might
interact with non uniform geometries, while observations of the CMB
have uncovered small scale temperature variations across space.
These models help us to grasp how these irregularities may be
connected to the fundamental structure of the universe and enable
exploration of diverse topologies and geometries beyond those
commonly assumed in cosmology. Koussour and Bennai \cite{17a}
studied Bianchi type-I cosmological models with a viscous bulk fluid
in the framework of $f(\mathcal{R,T})$ gravity. De et al \cite{17b}
explored the evolution using Bianchi type-I universe in
$f(\mathcal{Q})$ gravity. The effects of torsion on the evolution of
the Bianchi type-I universe in the context of $f(T)$ theory has been
studied in \cite{17c} ($T$ is the torsion). Solanke \cite{17d}
examined how locally rotationally symmetric Bianchi type-I cosmic
model behavior was influenced by modified $f(\mathcal{Q,T})$
gravity. The study of Bianchi type-I universe model with Noether
symmetry approach in energy-momentum squared gravity has been
studied in \cite{17e}.

The interesting inherent features of bouncing cosmology in
$\mathcal{MGT}s$ have garnered considerable attention from
researchers. Bajardi et al \cite{18} examined bouncing cosmology in
$f(\mathcal{Q})$ theory and computed the cosmic wave function using
Hamiltonian formalism. Mandal et al \cite{19} employed several
$f(\mathcal{Q})$ models to examine the stability of matter bounce.
The study of cosmic evolution by analyzing non singular bouncing
cosmological solutions in $f(\mathcal{R})$ theory has been explored
in \cite{20}. Ilyas et al \cite{21} examined the bouncing universe
using various $f(\mathcal{R})$ models. Bhardwaj et al \cite{22}
explored cosmic accelerated expansion by evaluating several
cosmographic parameters in this framework. Lohakare et al \cite{23}
assessed the influence and importance of the Gauss-Bonnet terms on
the evolution of the universe. Yousaf et al \cite{24} used energy
constraints to investigate the cosmic bounce phenomenon. Houndjo et
al \cite{25} investigated the universe evolution by applying a
reconstructed teleparallel model. Their findings suggested that
adding stiff matter to the model could achieve stability. The
influence of modified correction factors on the bouncing phenomenon
has been studied in \cite{25a}-\cite{25g}.

Bozza and Burni \cite{26} explored solutions to the anisotropy issue
in the framework of bouncing cosmology. Cai et al \cite{27}
developed a nonsingular cosmic bounce model using a scalar field,
successfully establishing a matter bounce scenario that addresses
the anisotropy challenge. Cai \cite{28} examined matter bounce
inflation scenarios through reconstructed models, which hold
potential for advancing observational cosmology in the future. Bamba
et al \cite{29} explored various scale factor models in
$f(\mathcal{R})$ theory to find viable bouncing solutions in this
$\mathcal{MGT}$. Research on bouncing cosmology within the modified
Gauss-Bonnet gravity framework has been documented in \cite{30}.
Amani \cite{31} analyzed bouncing solutions in $ f(\mathcal{R})$
gravity using the redshift parameter as a basis. Haro and Amoros
\cite{32} employed the Arnowitt-Deser-Misner formalism to construct
bounce solutions in $f(\mathcal{T})$ gravity, where $ \mathcal{T}$
denotes the torsion scalar. Additional studies have explored bounce
solutions within other extended gravity frameworks \cite{33}.

This literature encourages us to investigate the feasibility of a
non singular bounce solutions in the framework of $f(\mathcal{Q,C})$
gravity. The paper structure adheres to this framework. Section
\textbf{2} introduces the fundamentals of $f(\mathcal{Q,C})$ gravity
and analyzes various cosmological quantities to explore the
complexities of the universe. In section \textbf{3}, we analyze
several $f(\mathcal{Q,C})$ models to get the explicit field
equations which are helpful to comprehend the cosmic evolution.
Section \textbf{4} offers a detailed analysis of bouncing cosmology
and dynamics of redshift in this modified framework. The stability
analysis of cosmological bounce solutions is also examined in
section \textbf{5}. We demonstrate our main results in section
\textbf{6}.

\section{$f(\mathcal{Q,C})$ Theory and Bianchi type-I Model}

To analyze the cosmological properties of non-metric gravity, we
examine the most general formulation of affine connections as
\begin{equation}\label{1}
\Lambda^{\lambda}_{~\xi\eta}=\check{\Lambda}^{\lambda}_{~\xi\eta}
+\mathcal{K}^{\lambda}_{~\xi\eta}+{\mathcal{L}}^{\lambda}_{~\xi\eta},
\end{equation}
where $\mathcal{K}^{\lambda}_{~\xi\eta}$ is the contortion tensor, defined as
\begin{equation}\label{3}
\mathcal{K}^{\lambda}_{~\xi\eta}=\frac{1}{2}\mathrm{g}^{\lambda\upsilon}
(T_{\xi\upsilon\eta}+T_{\eta\upsilon\xi}+T_{\upsilon\xi\eta}).
\end{equation}
The Levi-Civita connection is expressed as
\begin{equation}\label{2}
\check{\Lambda}^{\lambda}_{\xi\eta}=\frac{1}{2}\mathrm{g}^{\lambda\upsilon}
(\partial_{\xi}\mathrm{g}_{\upsilon\eta}+\partial_{\eta}\mathrm{g}_{\upsilon\xi} -
\partial_{\upsilon}\mathrm{g}_{\xi\eta}).
\end{equation}
The disformation tensor and non-metricity are given by
\begin{eqnarray}\label{4}
{\mathcal{L}}^{\lambda}_{~\xi\eta}&=&\frac{1}{2}
\mathrm{g}^{\lambda\upsilon}
(-\mathcal{Q}_{\xi\upsilon\eta}-\mathcal{Q}_
{\eta\upsilon\xi}+\mathcal{Q}_{\upsilon\xi\eta}),
\\\label{5} \mathcal{Q}_{\varrho\xi\eta}&=&
\nabla_{\varrho} \mathrm{g}_{\xi\eta}=\partial_
{\varrho} \mathrm{g}_{\xi\eta}-\Lambda^{\upsilon}_{\varrho\xi}
\mathrm{g}_{\upsilon\eta}-\Lambda^{\upsilon}_
{\varrho\eta}\mathrm{g}_{\xi\upsilon}.
\end{eqnarray}
Consequently, the non-metricity scalar is characterized as
\begin{equation}\label{6}
\mathcal{Q}=-\frac{1}{4}\mathcal{Q}_{\gamma\upsilon\xi}
\mathcal{Q}^{\gamma\upsilon\xi}+\frac{1}{2}
\mathcal{Q}_{\gamma\upsilon\xi}\mathcal{Q}^{\upsilon\xi\gamma} +\frac{1}{4}
\mathcal{Q}_{\gamma}\mathcal{Q}^{\gamma}-
\frac{1}{2}\mathcal{Q}_{\gamma}\tilde{\mathcal{Q}}^{\gamma}.
\end{equation}
Applying the torsion-free and curvature-free constraints, we have
\begin{eqnarray}\label{7}
&&\check{\mathcal{R}}_{\xi\eta}+\check
{\nabla}_{\gamma} \mathcal{L}^{\gamma}_{~\xi\eta}-
\check{\nabla}_{\eta}\tilde{\mathcal{L}}_{\xi}+
\tilde{\mathcal{L}}_{\gamma}\mathcal{L}^{\gamma}_{~\xi\eta}-
\mathcal{L}_{\gamma\upsilon\eta}\mathcal{L}^{\gamma\upsilon}_{~\xi}=0,
\\\label{8}
&&\check{\mathcal{R}}+\check{\nabla}_{\gamma}
(\mathcal{L}^{\gamma} -\tilde{\mathcal{L}}^{\gamma})-\mathcal{Q}=0.
\end{eqnarray}
Every value denoted by $\check{}$  is computed in relation to the
Levi-Civita connection. Since $\mathcal{Q}^{\gamma}-\tilde
{\mathcal{Q}}^{\gamma}=\mathcal{L}^{\gamma}-\tilde{\mathcal{L}}^{\gamma}$,
we consider the boundary term as
\begin{equation}\label{9}
\mathcal{C}=\check{\mathcal{R}}-\mathcal{Q}=- \check{\nabla}_{\gamma}(\mathcal{Q}^{\gamma}-
\tilde{\mathcal{Q}}^{\gamma}) =-\frac{1}{\sqrt{-\mathrm{g}}}
\partial_{\gamma}[\sqrt{-\mathrm{g}}(\mathcal{Q}^{\gamma}
-\tilde{\mathcal{Q}}^{\gamma})].
\end{equation}
Now, the integral action of $f(\mathcal{Q},\mathcal{C})$ gravity is
expressed as \cite{13}
\begin{equation}\label{10}
S=\int\frac{1}{2\kappa}\left(f(\mathcal{Q},\mathcal{C})+
2\kappa\mathcal{L}_{m}\right)\sqrt{-\mathrm{g}}d^{4}x.
\end{equation}
In this context, $\mathrm{g}$ signifies the characteristic of the
metric tensor and $\mathcal{L}_{m}$ represents the
matter-Lagrangian. The corresponding field equations are
\begin{eqnarray}\nonumber
\mathcal{T}_{\xi\eta}&=& \frac{2}{\sqrt{-\mathrm{g}}}
\partial_{\lambda}({\sqrt{-\mathrm{g}}}
f_{\mathcal{Q}}\mathcal{P}^{\lambda}_{~\xi\eta})
+(\mathcal{P}_{\xi\gamma\upsilon}\mathcal{Q}_{\eta}^{~~\gamma\upsilon}-2\mathcal{P}_
{\gamma\upsilon\eta}\mathcal{Q}^{\gamma\upsilon}_{~~\xi})f_{\mathcal{Q}}
\\\label{11}
&-&\frac{f}{2}\mathrm{g}_{\xi\eta}+ (\frac{\mathcal{C}}{2}\mathrm{g}_{\xi\eta}
-\check{\nabla}_{\xi}\check{\nabla}_{\eta}+ \mathrm{g}_
{\xi\eta}\check{\nabla}^{\gamma}\check{\nabla}_{\gamma}
-2\mathcal{P}^{\lambda}_{~~\xi\eta}\partial_{\lambda})f_{\mathcal{C}},
\end{eqnarray}
where $f_{\mathcal{Q}}=\frac{\partial f}{\partial \mathcal{Q}}$,
$f_{\mathcal{C}} =\frac{\partial f} {\partial \mathcal{C}}$ and
$\mathcal{EMT}$ is denoted by $\mathcal{T}_{\xi\eta}$. It is
observed that the affine connection is independent of the metric
tensor. Thus, the connection field equations may be derived by
modifying the action concerning the affine connection as stated in
\cite{13}.

To study the dynamics of a bouncing universe, the Bianchi type-I
spacetime is considered as
\begin{equation}\label{19}
ds^{2}=-dt^{2}+\mathrm{a}^{2}dx^{2}+(dy^{2}+dz^{2})\mathrm{b}^{2}.
\end{equation}
The Friedmann Robertson Walker universe can be obtained under the
condition $a(t)=b(t)$. In this analysis, we introduce a valuable
constraint defined by the relation $a(t)=b(t)^{\Omega}$, which
enforces a constant ratio between the shear scalar and expansion
scalar. Here, $\Omega$ is a constant where $\Omega \neq 0,1$,
ensuring the presence of non trivial solutions. The constraint
$\mathrm{a}=\mathrm{b}^{\Omega}$ is chosen for the mathematical
convenience and physical significance in the context of Bianchi
type-I cosmology. This relationship simplifies the equations
governing the dynamics of the universe, allowing more tractable
solutions and meaningful physical interpretations. In the revised
manuscript, we have included the references and provide a detailed
discussion on the rationale behind choosing the constraint
$\mathrm{a}=\mathrm{b}^{\Omega}$ along with its physical
implications and contexts. The constraint
$\mathrm{a}=\mathrm{b}^{\Omega}$ reduces the number of free
parameters in the metric, making the equations of motion simpler and
more manageable. This is particularly useful in obtaining analytical
solutions or simplifying numerical computations.

Anisotropic models with such constraints can be tested against
observational data such as the cosmic microwave background and large
scale structure. The parameter $m$ can be related to physical
properties of the early universe, such as the nature of primordial
anisotropies or the influence of certain fields on the geometry of
the universe. By changing $m$, one can explore different regimes of
anisotropy and their physical implications. In the early universe,
where deviations from isotropy and homogeneity are more pronounced,
constraints like $\mathrm{a}=\mathrm{b}^{\Omega}$ can provide
insights into the initial conditions and the mechanisms driving the
evolution of anisotropies. This condition implies that the Hubble
cosmic expansion can achieve isotropy when the shear to expansion
ratio is constant \cite{34}. This approach has been utilized by
several researchers to derive cosmological solutions \cite{35}. By
maintaining a constant ratio of the shear scalar to the expansion
scalar, it becomes possible to simplify the equations and obtain
meaningful insights into the behavior of the universe. The constant
ratio condition helps in achieving isotropy in the cosmic expansion
that aligns with observational evidence of cosmos. Several studies
have employed this approach to get cosmic solutions \cite{36}.

We assume a configuration with perfect matter distribution as
\begin{equation}\nonumber
\mathcal{T}_{\xi\eta}=(\varrho+\mathcal{P}) u_{\xi}u_{\eta}
+\mathcal{P}\mathrm{g}_{\xi\eta}.
\end{equation}
Here, the four velocity, density and pressure are denoted by
$u_{\xi}$, $\varrho$ and $\mathcal{P}$, respectively. By considering
the case with vanishing affine connection, we have
\begin{eqnarray}\label{23}
\mathcal{G}_{\xi\eta}&=&-h_{\xi\eta} (3\mathcal{H}^{2}
+2\dot{\mathcal{H}})+3\mathcal{H}^{2}u_{\xi}u_{\eta},
\\\label{24}
\check{\mathcal{R}}&=&12\mathcal{H}^{2}+6\dot{\mathcal{H}},
\\\label{25}
\mathcal{Q}&=&6\mathcal{H}^{2},
\\\label{26}
\mathcal{C}&=&\check{\mathcal{R}}-\mathcal{Q}
=6(\mathcal{H}^{2}+\dot{\mathcal{H}}),
\end{eqnarray}
with $h_{\xi\eta}=\mathrm{g}_{\xi\eta} + u_{\xi}u_{\eta}$. The
resulting field equations are
\begin{eqnarray}\label{27}
\varrho&=&-\frac{1}{2} f(\mathcal{Q,C})+3 \big(\mathcal{H}^2+
\dot{\mathcal{H}}\big) f_\mathcal{C}-3 \mathcal{H}
\dot{f_\mathcal{C}}-4 \mathcal{H}^2 \Omega f_\mathcal{Q} -2
\mathcal{H}^2 f_\mathcal{Q},
\\\nonumber
\mathcal{P}&=&\frac{1}{2} f(\mathcal{Q,C})- 3 \big(\mathcal{H}^2
+\dot{\mathcal{H}}\big) f_\mathcal{C}+\ddot{f_\mathcal{C}} -2
\mathcal{H}^2 \Omega f_\mathcal{Q} +4 \mathcal{H}^2 f_\mathcal{Q}
\\\label{28}
&+&2 \mathcal{H} \dot{f_\mathcal{Q}}-2 \dot{\mathcal{H}}
f_\mathcal{Q}.
\end{eqnarray}

\section{Viable $f(\mathcal{Q,C})$ Models}

This research investigates the impact of several $f(\mathcal{Q,C})$
models on cosmic evolution. Our aim is to get a comprehensive grasp
of fundamental elements of astrophysics and theoretical cosmology.
The correction terms offered by this $\mathcal{MGT}$ present
significant potential to investigate novel physical discoveries.
Consequently, examining $f(\mathcal{Q,C})$ models is essential to
comprehend the processes that govern cosmic bounce occurrences. In
the following sections, we examine three particular models of this
theory.

\subsection*{Model 1}

First, we analyze the $f(\mathcal{Q,C})$ model with arbitrary
constants ($\mu_{1}$ and $\mu_{2}$) as \cite{49}
\begin{equation}\label{32}
f(\mathcal{Q,C})=\mu_{1}\mathcal{Q}+\mu_{2}\mathcal{C}^2.
\end{equation}
Numerous studies have explored this model due to its simplicity
\cite{39}. This framework is particularly advantageous for
addressing complex problems and gaining insights into fundamental
principles. It enables more accurate calculations and improves our
understanding of gravitational processes. Thus, this model provides
both theoretical and observational implications for understanding
the fundamental principles of gravitational physics \cite{40}. The
field equations \eqref{27} and \eqref{28} corresponding to this
model become
\begin{eqnarray}\nonumber
\varrho&=&(\Omega+2)^2 \big(2 (\Omega-1) (\Omega+11) \mu _2
\dot{\mathrm{b}}^4 -12 (\Omega-4) \mu _2
\mathrm{b}\dot{\mathrm{b}}^2 \mathrm{b}'') -\mathrm{b}^2 \big((4
\Omega+5)
\\\label{33}
&\times&\mu _1 \dot {\mathrm{b}}^2 -18 \mu _2
\big(\ddot{\mathrm{b}}^2 -2\dddot{\mathrm{b}}
\dot{\mathrm{b}}\big)\big)\big)(9 \mathrm{b}^4)^{-1},
\\\nonumber
\mathcal{P}&=&(-(\Omega+2) \big(2 \big(\Omega^3-39 \Omega+ 38\big)
\mu _2 \dot{\mathrm{b}}^4+ 6 \mathrm{b}^3 \big(\mu _1
\ddot{\mathrm{b}}-6 \mu _2 \ddddot{\mathrm{b}}\big)+12 (\Omega
\\\nonumber
&\times&(\Omega+11)-18) \mu _ 2 \mathrm{b} \dot{\mathrm{b}}^2
\ddot{\mathrm{b}} + \mathrm{b}^2\big((\Omega-4) (2 \Omega+5) \mu _1
\dot{\mathrm{b}}^2+6 \mu _2 \big(-(\Omega
\\\label{34}
&-&16) \ddot{\mathrm{b}}^2-4 (\Omega-4) \dddot{\mathrm{b}}
\dot{\mathrm{b}}\big)\big)\big))(9 \mathrm{b}^4)^{-1}.
\end{eqnarray}

\subsection*{Model 2}

We consider another functional of $f(\mathcal{Q},\mathcal{C})$ as
\cite{49}
\begin{equation}\label{35}
f(\mathcal{Q,C})=\mu_{3} \mathcal{Q}^{\lambda +1}
+\mu_{2}\mathcal{C}^2,
\end{equation}
where $\lambda$ and $\mu_{3}$ are arbitrary constants.
The outcomes of this model are crucial for understanding
the dynamics of universe \cite{41}. Using Eq.\eqref{35} in \eqref{27} and \eqref{28}, we have
\begin{eqnarray}\nonumber
\varrho&=&2(\Omega+2)^2 \mu _2 \big(9 \mathrm{b}^2
\ddot{\mathrm{b}}^2+(\Omega-1) (\Omega+11) \dot{\mathrm{b}}^4-18
\mathrm{b}^2 \dddot{\mathrm{b}} \dot{\mathrm{b}}-6 (\Omega-4)
\mathrm{b} \dot{\mathrm{b}}^2 \ddot{\mathrm{b}} \big)
\\\label{36}
&\times&(9 \mathrm{b}^4)^{-1}-2^{\lambda } 3^{-\lambda -2} \mu _3 (2
\lambda +4 (\lambda +1)\Omega+5) \big(\frac{(\Omega+2)^2
\dot{\mathrm{b}}^2}{\mathrm{b}^2}\big)^{\lambda +1},
\\\nonumber
\mathcal{P}&=&-3^{-\lambda -2} (\Omega+2) \big(2\ 3^{\lambda } \mu
_2 \big(\big(\Omega^3-39 \Omega+38\big)\dot{\mathrm{b}}^4 -3
\mathrm{b}^2 \big(6 \mathrm{b} \ddddot{\mathrm{b}}+(\Omega-16)
\\\nonumber
&\times& \ddot{\mathrm{b}}^2\big)-12 (\Omega-4)\mathrm{b}^2
\dddot{\mathrm{b}} \dot{\mathrm{b}} +6 (\Omega (\Omega+11)-18)
\mathrm{b} \dot{\mathrm{b}} ^2 \ddot{\mathrm{b}}\big)-2^{\lambda }
\mu _3 \mathrm{b}^2
\\\nonumber
&\times&\big(\frac{(\Omega+2)^2 \dot{\mathrm{b}}^2}
{\mathrm{b}^2}\big) ^{\lambda } \big(6 \big(2 \lambda ^2+\lambda
-1\big) \mathrm{b} \ddot{\mathrm{b}}+\big(2 \big(-6 \lambda
^2+\lambda +10\big)-2 (\lambda +1)
\\\label{37}
&\times&\Omega^2+3 \Omega\big)
\dot{\mathrm{b}}^2\big)\big)(\mathrm{b}^4)^{-1}.
\end{eqnarray}

\subsection*{Model 3}

Here, we assume functional form with arbitrary constant $\mu_{4}$ as
\cite{49a}
\begin{equation}\label{38}
f(\mathcal{Q,C}) =
\mathcal{Q}+\frac{\mu_{4}}{\mathcal{Q}}+\mu_{2}\mathcal{C}^{2}.
\end{equation}
This functional form provides a robust framework for comprehending
the universe behavior during the bouncing phase. The phrase
$\frac{\mu_{4}}{\mathcal{Q}}$ denotes an inverse curvature,
essential for precisely depicting the effects of high curvature
regimes pertinent throughout the bounce. This alteration guarantees
that the model maintains its physical significance during various
periods of the cosmic evolution especially when shifting from
contraction to expansion. Moreover, the existence of the
$\mu_{4}\mathcal{C}^{2}$ component signifies the effect of field
perturbations on cosmic phenomena. The field equations for this
model are
\begin{eqnarray}\nonumber
\varrho&=&8 (\Omega-1) (\Omega+2)^4 (\Omega+11) \mu _2
\dot{\mathrm{b}}^6-48 (\Omega-4)(\Omega+2)^4 \mu _2 \mathrm{b}
\dot{\mathrm{b}}^4 \ddot{\mathrm{b}}-4 (\Omega
\\\nonumber
&+&2) ^4 \mathrm{b}^2  \dot{\mathrm{b}}^2\big(-18 \mu
_2\ddot{\mathrm{b}}^2+(4 \Omega+5) \dot{\mathrm{b}}^2+ 36 \mu _2
\dddot{\mathrm{b}} \dot{\mathrm{b}}\big) +9 (4 \Omega-1) \mu _4
\mathrm{b}^6 (36
\\\label{39}
&\times&(\Omega+2)^2 \mathrm{b}^4 \dot{\mathrm{b}}^2)^{-1},
\\\nonumber
\mathcal{P}&=&-(-270 \mu _4 \mathrm{b}^7 \ddot{\mathrm{b}}+8
(\Omega+2)^4 \big(\Omega^3-39 \Omega +38\big) \mu _2
\dot{\mathrm{b}}^8 -9 (\Omega (2 \Omega+3)
\\\nonumber
&-&32) \mu _4 \mathrm{b}^6 \dot{\mathrm{b}}^2+48 (\Omega+2)^4
(\Omega (\Omega+11)-18) \mu _2 \mathrm{b} \dot{\mathrm{b}}^6
\dot{\mathrm{b}}+24 (\Omega+2)^4 \mathrm{b}^3 \dot{\mathrm{b}}^4
\\\nonumber
&\times&\big(\dot{\mathrm{b}}-6 \mu _2 \ddddot{\mathrm{b}}\big)+4
(\Omega+2)^4 \mathrm{b}^2 \dot{\mathrm{b}}^4 \big(-6 (\Omega-16) \mu
_2 \dot{\mathrm{b}}^2 +(\Omega-4) (2 \Omega
\\\label{40}
&+&5) \dot{\mathrm{b}}^2-24 (\Omega-4) \mu _2 \dddot{\mathrm{b}}
\dot{\mathrm{b}}\big)) (36 (\Omega+2)^3 \mathrm{b}^4
\dot{\mathrm{b}}^4)^{-1}.
\end{eqnarray}

\section{Analysis of Bouncing Cosmology}

This section explores the viability of different bouncing
cosmological models, including the symmetric bounce, oscillatory
bounce, matter bounce and exponential bounce. Each of these models
is analyzed to describe the cosmic evolution. Through a comparative
analysis, we aim to highlight their strengths, limitations and
overall relevance to modern cosmological theories.

Bouncing cosmological solutions give a compelling alternative to the
standard cosmological models by addressing the challenge of initial
singularities like the big bang. Unlike standard frameworks where
the universe begins from a singular point, bouncing models enable a
smooth transition from contraction to expansion, avoiding the
initial singular state. This mechanism not only redefines our
understanding of the universe birth but also allows for a novel
examination of the early universe phenomena. Thus, bouncing
cosmological models challenge the limitations of singularity based
models by satisfying specific conditions that maintain stability and
coherence in cosmic evolution. These criteria are crucial for
ensuring the model viability, opening the door to further
exploration of non-singular cosmic histories and a deep grasp of the
cosmic foundational processes. For realistic $f(\mathcal{Q},
\mathcal{C})$ gravity, the following conditions should be satisfied
in order to construct the realistic bouncing pattern of the
universe.
\begin{itemize}
\item
The decreasing nature of the scale factor indicates that the
universe is in a phase of contraction, while its increasing behavior
signifies the era of cosmic expansion. If the bouncing model is
non-singular then the scale factor should be minimum close to the
bouncing spot.
\item
The universe experiences contraction when Hubble parameter is
negative and cosmos undergoes expansion when Hubble parameter is
positive. The Hubble parameter vanishes at the bouncing point.
\item
The $\mathcal{EOS}$ parameter depicts that the universe is passing
through a phantom phase when $\omega<-1$ and quintessence phase when
$-1<\omega < -1$.
\item
The energy density must be positive, finite as well as maximum and
pressure should be negative for the existence of non singular
bounce.
\end{itemize}
These physical characteristics are important to determine the viable
cosmological bounce solutions.

\subsection{Evolution of the Scale Factor}

\begin{figure}\center
\epsfig{file=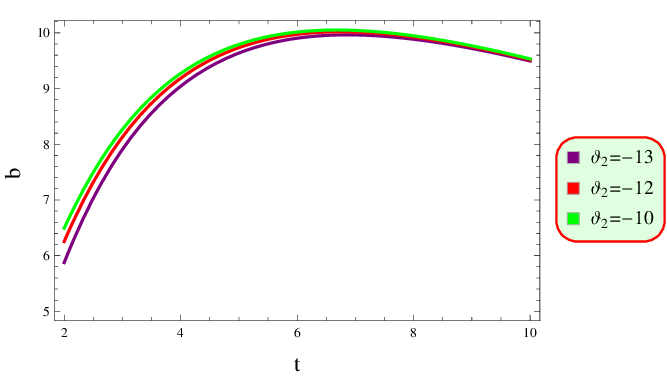,width=.5\linewidth} \caption{Evolution of the
scale factor across various $\vartheta_{2}$ values.}
\end{figure}

The scale factor is essential for studying the behavior of the
universe that could be expanding, contracting or even going through
a cyclic bounce. The positive time dependent function indicates the
universe expansion and relative size at any point in time.
Examination of the scale factor provides valuable insights into the
evolution of the cosmos. To construct a bouncing cosmological model,
we define a scale factor as \cite{37}
\begin{equation}\label{29}
\mathrm{b}=\beta e^{\frac{\vartheta _3 t^{n+1}}{n+1}
+\frac{\vartheta _2 t^n}{n}+\vartheta _1 t},
\end{equation}
where $\vartheta_{1}$, $\vartheta_{2}$, $\vartheta_{3}$ and $n$
represent non zero arbitrary constants with $\beta$ denotes the
integration constant. The considered representation of the scale
factor is determined by various significant physical and
mathematical factors in cosmology. This form enables a broad
description of the scale factor evolution and covering a variety of
viable cosmological behaviors. This form of the scale factor
represents several phases of the cosmic expansion, including early
time inflation, radiation domination, matter domination and late
time accelerated expansion. Singh et al \cite{37} employed this
parametric representation of the scale factor to evaluate the
phenomenon of bouncing cosmology in $f(\mathcal{R,T})$ framework.
Motivated by their work, we examine the behavior of cosmic
parameters in the framework of $f(\mathcal{Q,C})$ theory using the
same parametric form of the scale factor. Figure \textbf{1} shows
that the scale factor is finite and positive across the entire
cosmic timeline.

\subsection{Analysis of Hubble Parameter}

\begin{figure}
\epsfig{file=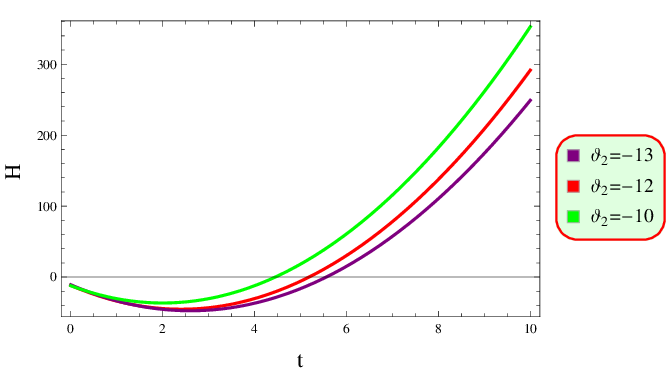,width=.5\linewidth}
\epsfig{file=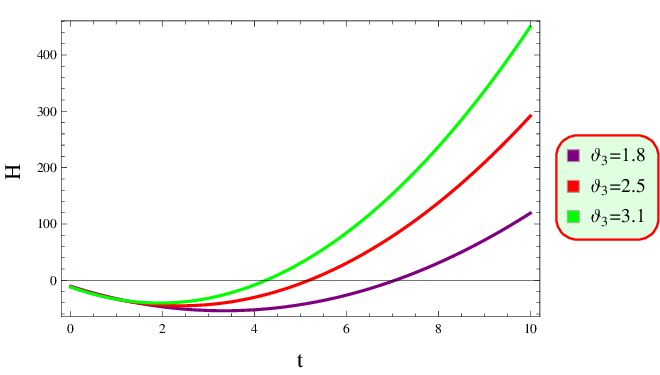,width=.5\linewidth}\center
\epsfig{file=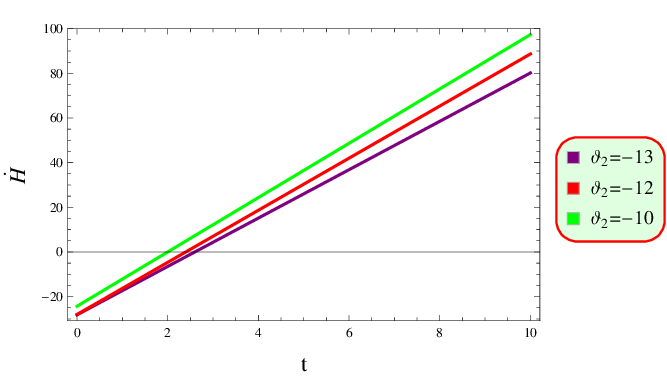,width=.5\linewidth}\caption{Plots of the Hubble
parameter and its temporal derivative for different parameter
values.}
\end{figure}

It is a fundamental cosmological quantity used to analyze
cosmic evolution and investigate various applications
in cosmology. The amount of the Hubble parameter is often
treated as a parameter in theoretical models to explore
different cosmic phenomena \cite{38}. It is defined as
\begin{equation}\label{30}
\mathcal{H}=\frac{\dot{\mathrm{b}} \Omega+2 \dot{\mathrm{b}}}{3
\mathrm{b}}.
\end{equation}
The Hubble parameter and its time derivative play a key role in
determining energy density and pressure in cosmological models.
However, in this scenario, field equations cannot be fully resolved
without certain constraints. We explore a scenario that places a
limit on the Hubble parameter, focusing on a case that includes a
cosmological bounce. The units used in this specific model reflect
cosmic parameters, providing clarity in depicting cosmic evolution
on a grand scale. Figure \textbf{2} demonstrates a transition in the
Hubble parameter from a contraction phase $(\mathcal{H}<0)$ to an
expansion phase $(\mathcal{H}>0)$ at the bounce point, where
$\mathcal{H}=0$. Additionally, it determines how the value of
$\Omega$ affects the occurrence of the bounce as increasing $\Omega$
leads to a later bouncing time. Table \textbf{1} delineates the
attributes of the Hubble parameter for varying amounts of $\Omega$
and $\vartheta_2$, while maintaining $\vartheta_3$ constant at
$2.5$. Table \textbf{2} delineates the characteristics of the Hubble
parameter for different values of $\Omega$ and $\vartheta_3$, with
$\vartheta_2$ fixed at $-12$. Figure \textbf{2} demonstrates the
temporal derivative of the Hubble parameter, highlighting the shift
between cosmic contraction and expansion phases.
\begin{table}\caption{Impact of $\vartheta_{2}$
on the Hubble parameter keeping $\vartheta_{3}$ fixed.}
\begin{center}
\begin{tabular}{|c|c|c|c|c|}
\hline $\vartheta_{2}$ & $\vartheta_{3}$   & $\Omega$    & Time & Nature of H
\\
\hline -13     & 2.5        & 4.5   & $0<t<5.87$   & Contraction
\\
\hline -12   & 2.5          & 5       & $0<t<5.13$   & Contraction
\\
\hline -10    & 2.5          & 5.3  & $0<t<4.47$   & Contraction
\\
\hline -13     & 2.5         & 4.5    & $5.87<t<\infty$  & Expansion
\\
\hline -12   & 2.5            & 5    &  $5.13<t<\infty$ & Expansion
\\
\hline  -10   & 2.5           & 5.3  & $4.47<t<\infty$ & Expansion
\\
\hline
\end{tabular}
\end{center}
\end{table}
\begin{table}\caption{Impact of $\vartheta_{3}$ on the Hubble parameter
keeping $\vartheta_{2}$ fixed.}
\begin{center}
\begin{tabular}{|c|c|c|c|c|}
\hline $\vartheta_{2}$ & $\vartheta_{3}$ & $\Omega$  &Time & Nature of H
\\
\hline -12   & 1.8  &4.5   & $0<t<7.13$   & Contraction
\\
\hline -12   & 2.5   &5   & $0<t<5.24$   & Contraction
\\
\hline -12   & 3.1  &5.3    & $0<t<4.31$   & Contraction
\\
\hline -12   & 1.8  &4.5 & $7.13<t<\infty$  & Expansion
\\
\hline -12   & 2.5  &5 &  $5.24<t<\infty$ & Expansion
\\
\hline -12   & 3.1  &5.3 & $4.31<t<\infty$ & Expansion
\\
\hline
\end{tabular}
\end{center}
\end{table}

\subsection{Investigation of Rate of Cosmic Expansion}

\begin{figure}\center
\epsfig{file=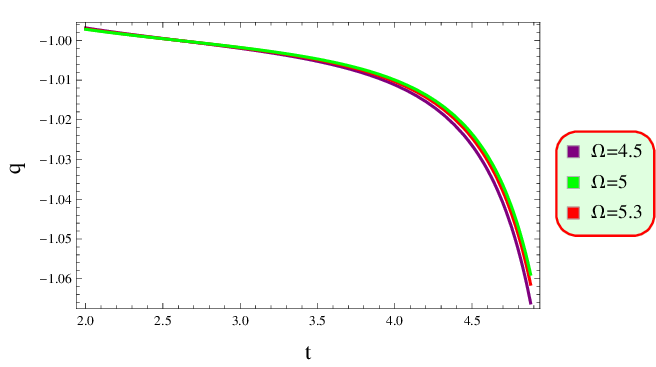,width=.5\linewidth} \caption{Graph of
deceleration parameter for various values of $\Omega$.}
\end{figure}
The deceleration parameter $(q)$ is given by
\begin{equation}\label{31}
q=-\dot{\mathcal{H}}\mathcal{H}^{-2}-1.
\end{equation}
This parameter reflects the rate of cosmic expansion, such as
positive values of $q$ signify a decelerated universe, whereas the
negative values of $q$ represent the accelerated cosmic model.
Figure 3 shows that the decelerated parameter consistently retains
both contraction and expansion stages.

\subsection{Analysis of Matter Variables}

The visual representation of energy density and pressure in bouncing
cosmology offers crucial insights into the characteristics of
bounce. These matter variables elucidate the dynamics of the
universe during the bouncing period and the viability of alternative
bouncing scenarios in a theoretical framework. The graphical
analysis in Figures \textbf{4}-\textbf{6} reveal that the behavior
of energy density is positive, whereas the pressure is negative
before and after the bounce. These graphical behavior align with the
characteristics of $\mathcal{DE}$, suggesting accelerated expansion.
The distinctive properties of energy density and pressure allow for
assessing the viability of the bouncing model, underscoring its
importance in advancing our understanding of cosmic evolution.
\begin{figure}
\epsfig{file=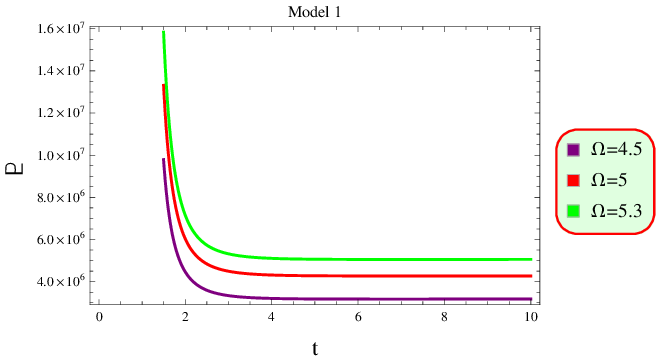,width=.5\linewidth}
\epsfig{file=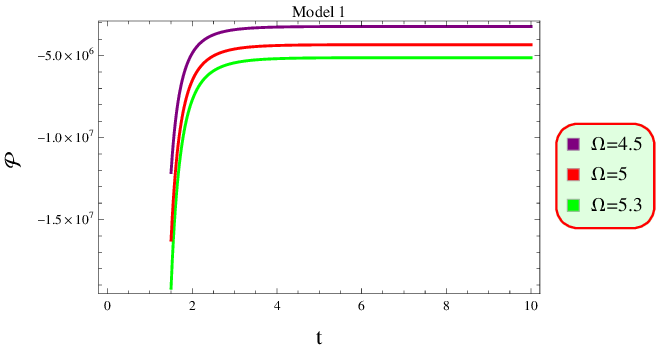,width=.5\linewidth} \caption{Behavior of matter
variables corresponding to Model \textbf{1}.}
\end{figure}
\begin{figure}
\epsfig{file=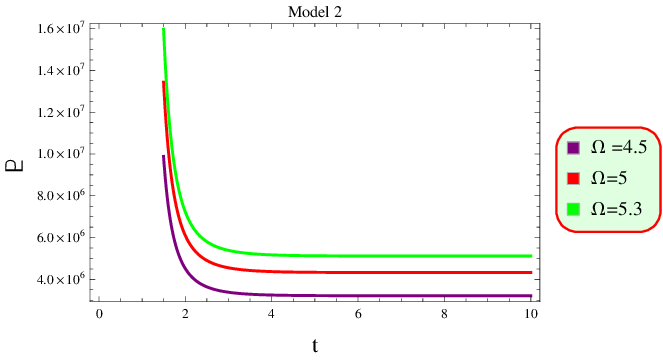,width=.5\linewidth}
\epsfig{file=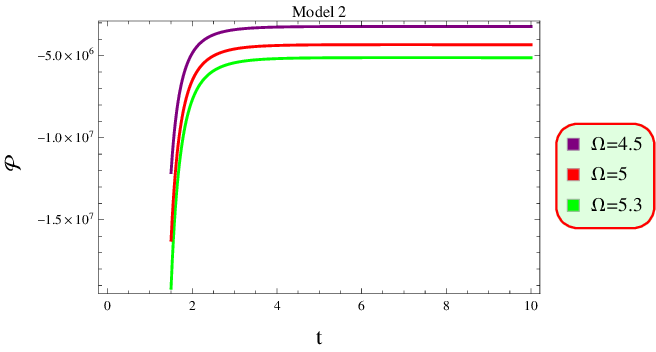,width=.5\linewidth} \caption{Plots of fluid
parameters for Model \textbf{2}.}
\end{figure}
\begin{figure}
\epsfig{file=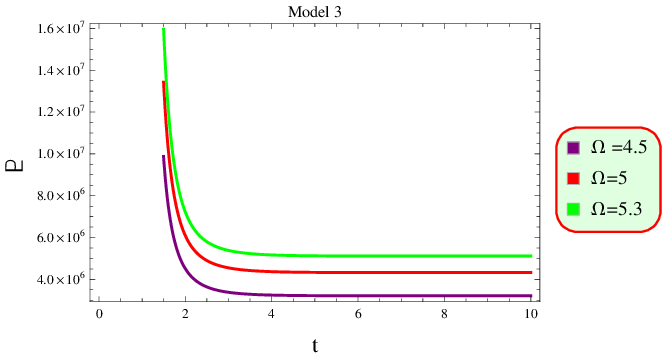,width=.5\linewidth}
\epsfig{file=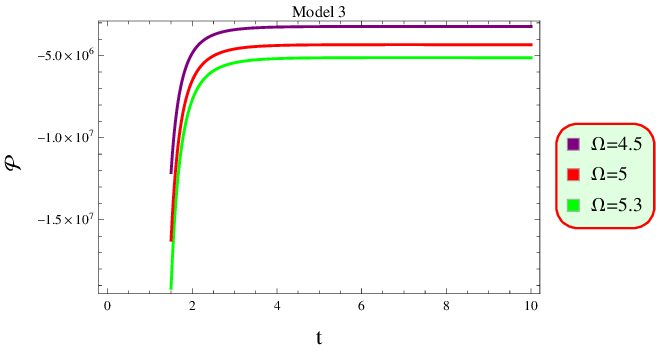,width=.5\linewidth} \caption{Evolution of
matter contents with respect to model \textbf{3}.}
\end{figure}

\subsection{Analysis of State Parameter}

The $\mathcal{EOS}$ parameter defined as
$\omega=\frac{\mathcal{P}}{\varrho}$ can be categorized according to
different phases of cosmic evolution. The eras characterized by
matter, namely dust radiative fluids and stiff matter are denoted by
$\omega = 0$, $\frac{1}{3}$ and $1$, respectively. The cosmic phases
of vacuum, phantom, and quintessence correspond to $\omega = -1$,
$\omega < -1$ and $-1 < \omega < -\frac{1}{3}$, respectively
\cite{42}. By using Eqs.\eqref{33} and \eqref{34}, we obtain the
$\mathcal{EOS}$ parameter for model \textbf{1} as
\begin{eqnarray}\nonumber
\omega&=&(\mu _2 \big(24 \big(\mathcal{H}^{(3)}
+2 \mathcal{H} \ddot{\mathcal{H}}+2 \dot{\mathcal{H}}^2\big)-36
\big(\dot{\mathcal{H}}+\mathcal{H}^2\big)^2\big)
-2 \mu _1 \big(2 \dot{\mathcal{H}}+(2 \Omega-7)
\\\nonumber
&\times& \mathcal{H}^2\big))(2 \big(-36 \mu _2
\mathcal{H} \ddot{\mathcal{H}}-\mathcal{H}^2 \big(36 \mu _2
\dot{\mathcal{H}}+(4 \Omega+5) \mu _1\big)+18 \mu _2 \dot{\mathcal{H}}^2+18 \mu _2
\\\label{41}
&\times& \mathcal{H}^4\big))^{-1}.
\end{eqnarray}
Similarly, for model \textbf{2}, we have
\begin{eqnarray}\nonumber
\omega&=&(6 \mu _2 \big(2 \mathcal{H}^{(3)}+4
\mathcal{H} \ddot{\mathcal{H}}-6 \mathcal{H}^2
\dot{\mathcal{H}}+\dot{\mathcal{H}}^2-3 \mathcal{H}^4
\big)-6^{\lambda } \mu _3 \big(\mathcal{H}^2\big)^{\lambda }
\big(\mathcal{H}^2 (-4 \lambda
\\\nonumber
&+&2 (\lambda +1) \Omega-7)-2\big(2 \lambda ^2+\lambda
-1\big) \dot{\mathcal{H}}\big))(18 \mu _2
\big(\big(\mathcal{H}^2-\dot{\mathcal{H}}\big)^2-2
\mathcal{H} \ddot{\mathcal{H}}\big)
\\\label{42}
&-&6^{\lambda } \mu _3 (2 \lambda +4 (\lambda +1)
\Omega+5) \big(\mathcal{H}^2\big)^{\lambda +1})^{-1}.
\end{eqnarray}
The value of the $\mathcal{EOS}$ parameter for model \textbf{3} is
\begin{eqnarray}\nonumber
\omega&=&(864 \mu _2 \mathcal{H}^5 \ddot{\mathcal{H}}-
36 \mathcal{H}^6 \big(36 \mu _2 \dot{\mathcal{H}}+2
\Omega-7\big) +10 \mu _4 \dot{\mathcal{H}}+72 \mathcal{H}^4 \big(6 \mu _2 \mathcal{H}^{(3)}
\\\nonumber
&+&\dot{\mathcal{H}} \big(3 \mu _2 \dot{\mathcal{H}}-
1\big)\big)+(2\Omega-1) \mu _4 \mathcal{H}^2-648 \mu _2
\mathcal{H}^8)(\mathcal{H}^2 \big(- 1296 \mu _2 \mathcal{H}^3 \ddot{\mathcal{H}}
\\\nonumber
&-&36 \mathcal{H}^4 \big(36 \mu _2 \dot{\mathcal{H}}
+4 \Omega+5\big)+648 \mu _2 \mathcal{H}^2
\dot{\mathcal{H}}^2+648 \mu _2 \mathcal{H}^6 +(4\Omega-1)
\\\label{43}
&\times&\mu _4\big))^{-1}.
\end{eqnarray}
Figure \textbf{7} shows the evolution of the $\mathcal{EOS}$
parameter for all models $\textbf{1}-\textbf{3}$. The
$\mathcal{EOS}$ parameter has a value of $\omega$ as $-1.03\pm
0.03$, which aligns well with the results reported by the Planck
collaboration \cite{43}. This behavior meets the conditions for the
quintessence regime $(-1 < \omega < -\frac{1}{3})$, aligning with
observable features of the universe.
\begin{figure}
\epsfig{file=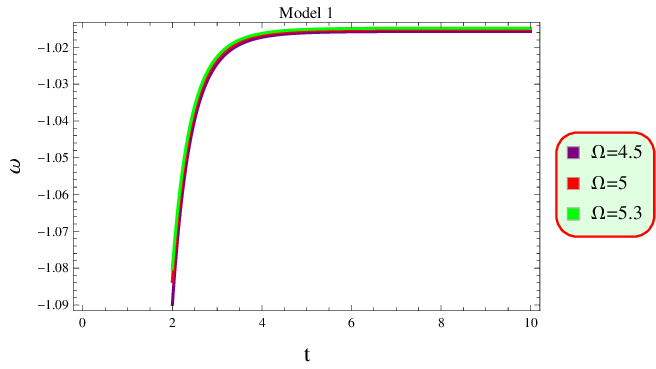,width=.5\linewidth}
\epsfig{file=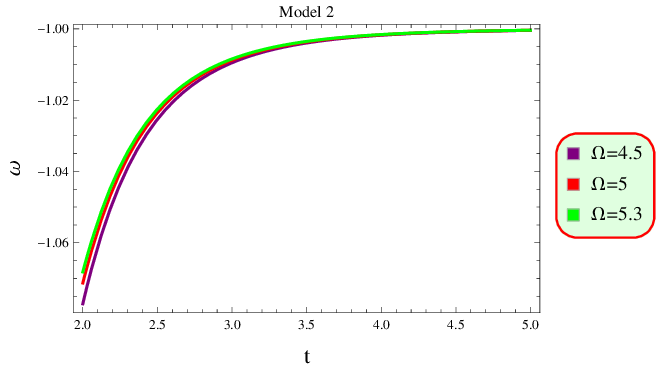,width=.5\linewidth}
\center\epsfig{file=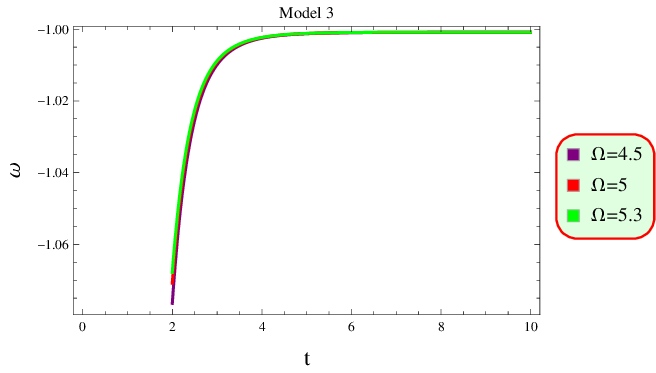,width=.5\linewidth}\caption{Behavior of
$\mathcal{EOS}$ parameter.}
\end{figure}

\subsection{Exploration of Energy Bounds}

These constraints are derived from the $\mathcal{EMT}$ that help to
evaluate the physical coherence of cosmological theories. Through
the implementation of these limitations, researchers investigate the
feasibility of various cosmic configurations. The $\mathcal{EC}s$
are categorized as follows, i.e., null ($0 \leq \varrho +
\mathcal{P}$), dominant ($0 \leq \varrho - \mathcal{P}$), weak ($0
\leq \varrho$ and $0 \leq \varrho + \mathcal{P}$) and strong ($0
\leq \varrho + \mathcal{P}$ and $0 \leq \varrho + 3\mathcal{P}$)
$\mathcal{EC}s$. This section presents a graphical analysis of the
energy constraints for each $f\mathcal{(Q,C)}$ model. The violation
of the $\mathcal{NEC}$ leads to the violation of all other
$\mathcal{EC}s$, illustrating the existence of a non-singular
bouncing universe. Figures \textbf{8}-\textbf{10} demonstrate the
presence of a non singular bounce model in this theoretical
framework.
\begin{figure}
\epsfig{file=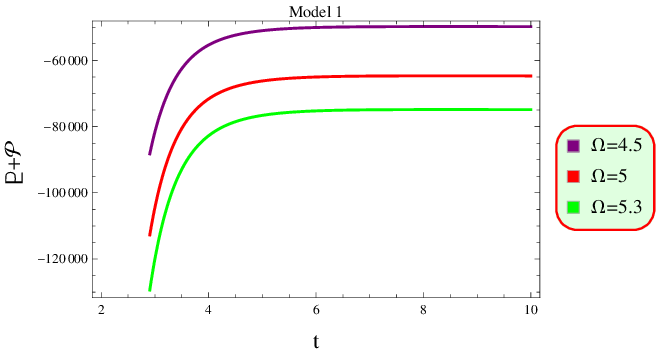,width=.5\linewidth}
\epsfig{file=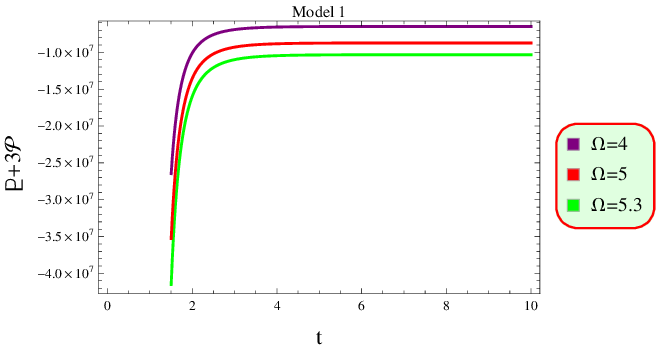,width=.5\linewidth}\center
\epsfig{file=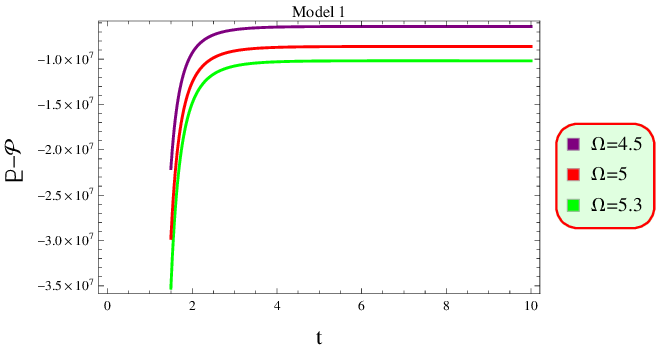,width=.5\linewidth}\caption{Graphs of energy
conditions corresponding to different parameter values for model
\textbf{1} across .}
\end{figure}
\begin{figure}
\epsfig{file=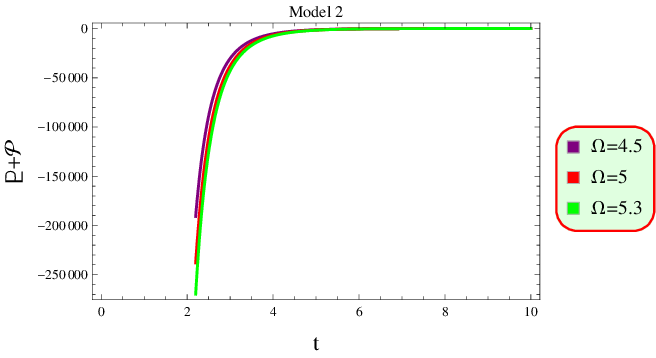,width=.5\linewidth}
\epsfig{file=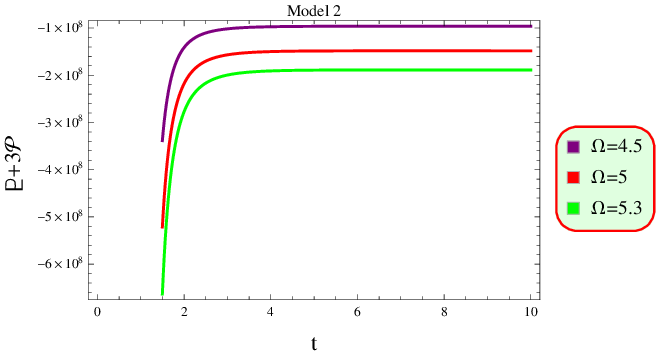,width=.5\linewidth}\center
\epsfig{file=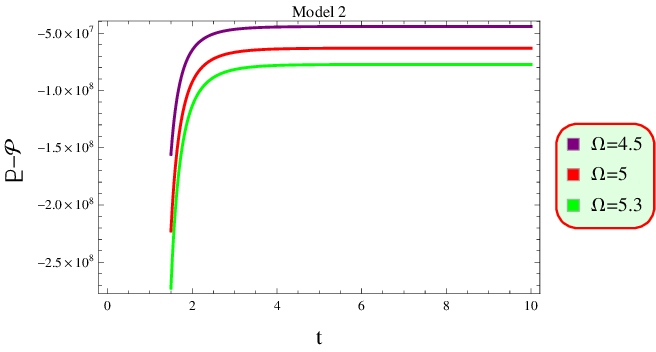,width=.5\linewidth}\caption{Plots of the energy
bounds for model \textbf{2}.}
\end{figure}
\begin{figure}
\epsfig{file=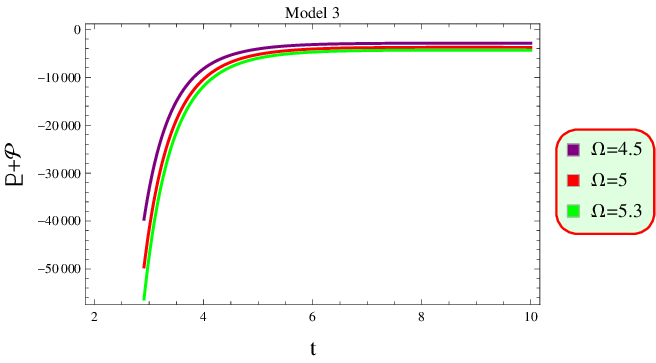,width=.5\linewidth}
\epsfig{file=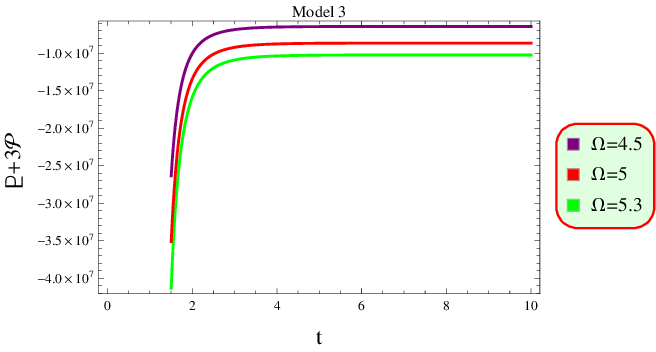,width=.5\linewidth}
\center\epsfig{file=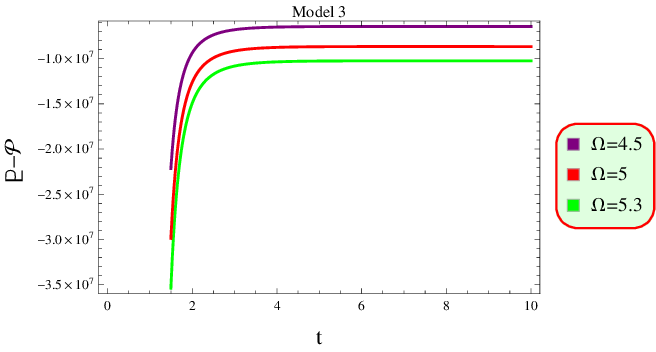,width=.5\linewidth}\caption{Behavior of
the energy conditions with respect to model \textbf{3}.}
\end{figure}

\subsection{Evaluation of Hubble Radius}

The Hubble radius (Hubble length) represents the visible distance
scale of the universe at a certain instant in time. Mathematically,
it is expressed as \cite{45}
\begin{equation}\label{49}
R_{H}=\frac{1}{\mathcal{H}}.
\end{equation}
Using Eqs.\eqref{29} and \eqref{30} in \eqref{49}, We derive the Hubble radius as
\begin{equation}\label{50}
R_{H}=\frac{1}{3}(\Omega+2)(\vartheta_{1}+\vartheta_{3}t^{n}+\vartheta_{2}t).
\end{equation}
The graphical representation of the Hubble radius can be observed in
Figure \textbf{11}. Prior to the bounce, cosmic time has a negative
value and remains relatively small which accentuates the influence
of the Hubble radius behavior. The behavior of $R_{H}$ demonstrates
a smooth shift from contraction to expansion, reflecting its inverse
relationship with the cumulative terms in the Hubble parameter.
\begin{figure}\center
\epsfig{file=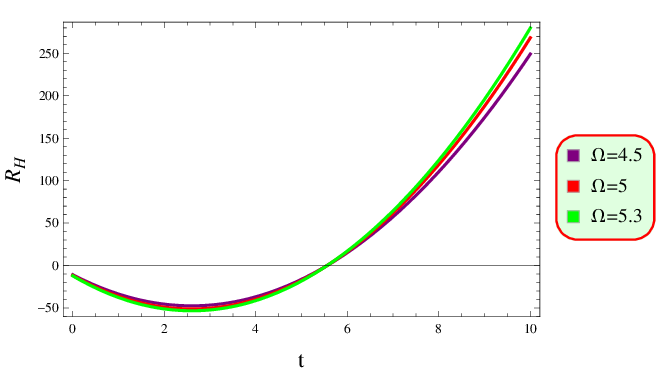,width=.5\linewidth}\caption{Behavior of the
Hubble radius $R_{H}$ as a function of cosmic time.}
\end{figure}

\subsection{Dynamics of Redshift}

The redshift parameter $(z)$ is used to explore the matter
configuration. A scale factor is defined as $\mathrm{b}(t) =
\mathrm{b}_{0}t^{\Upsilon}$, where $\Upsilon$ is a positive
arbitrary constant with a current value of $1$ \cite{46}. The
deceleration parameter can be represented as
\begin{equation}\label{53}
q=-1-\frac{\mathrm{b}\ddot{\mathrm{b}}}{\dot{\mathrm{b}}^{2}}=-1+\frac{1}{\Upsilon}.
\end{equation}
By changing the value of $\Upsilon$, we have
\begin{equation}\label{54}
\mathrm{b}(t)=t^{\frac{1}{1+q}} ,
\end{equation}
where  $q =-0.831_{-0.091}^{+0.091}$. The corresponding values of
$\mathcal{H}$ and $\mathcal{H}_{0}$ become
\begin{equation}\label{55}
\mathcal{H}=(1+q)^{-1}t^{-1},
\quad\mathcal{H}_{0}=(1+q)^{-1}t_{0}^{-1}.
\end{equation}
This indicates that $q$ and $\mathcal{H}_{0}$ are responsible for
the expansion of the cosmos. As a result of computing the
relationship between the scale factor and the redshift parameter, we
obtain
\begin{equation}\label{56}
\mathcal{H}=\mathcal{H}_{0}(1+z)^{1+q},
\quad\dot{\mathcal{H}}=-\mathcal{H}_{0}(1+z)^{2+2q}.
\end{equation}
The obtained value of non-metricity is
\begin{equation}\\\label{57}
\mathcal{Q}=6\mathcal{H}^{2}_{0}(1+z)^{2+2q}.
\end{equation}
The field equations in relation to the redshift function for model
\textbf{1} are
\begin{eqnarray}\nonumber
\varrho&=& \mathcal{H}_0^2 (z+1)^{2 q+2} \big(18
\big(\mathcal{H}_0-1\big) \big(\mathcal{H}_0+3\big) \mu _2
(z+1)^{2 q+2}-(4 \Omega+5)
\\\label{58}
&\times&\mu _1\big),
\\\nonumber
\mathcal{P}&=&\mathcal{H}_0 (z+1)^{2 q+2} \big(\mu _1 \big(\mathcal{H}_0 (7-2 \Omega)+2\big)-18
\big(\mathcal{H}_0-1\big) \big(\big(\mathcal{H}_0-1\big) \mathcal{H}_0
\\\label{59}
&-&4\big) \mu _2 (z+1)^{2 q+2}\big).
\end{eqnarray}
For the model \textbf{2}, we obtain
\begin{eqnarray}\nonumber
\varrho&=&\frac{1}{2} \big(36 \big(\mathcal{H}_0-1\big)
\mathcal{H}_0^2 \big(\mathcal{H}_0+3\big) \mu _2 (z+1)^{4
q+4}-2^{\lambda +1}3^{\lambda } \mu _3 (2 \lambda +4 (\lambda +1)
\\\label{60}
&\times&\Omega+5) \big(\mathcal{H}_0^2 (z+1)^{2 q+2}\big){}^{\lambda +1}\big),
\\\nonumber
\mathcal{P}&=&\mathcal{H}_0 (z+1)^{2 q+2} \big(-6^{\lambda }
\mu _3 \big(2 (\lambda +1) (2 \lambda
-1)+\mathcal{H}_0 (2 \lambda  (\Omega-2)+2 \Omega-7)\big)
\\\nonumber
&\times&\big(\mathcal{H}_0^2 (z+1)^{2 q+2} \big){} ^{\lambda }-18
\big(\mathcal{H}_0-1\big)
\big(\big(\mathcal{H}_0-1\big) \mathcal{H}_0 -4\big) \mu _2
\\\label{61}
&\times& (z+1)^{2 q+2}\big).
\end{eqnarray}
Similarly, the field equations for the model \textbf{3} give
\begin{eqnarray}\nonumber
\varrho&=&\frac{(4\Omega-1) \mu _4 (z+1)^{-2 (q+1)}}
{36 \mathcal{H}_0^2}-\mathcal{H}_0^2 (4\Omega+5) (z+1)^{2
q+2}+18 \big(\mathcal{H}_0-1\big)
\\\label{62}
&\times& \mathcal{H}_0^2 \big(\mathcal{H}_0+3\big) \mu _2 (z+1)^{4 q+4},
\\\nonumber
\mathcal{P}&=&\frac{\mu _4 \big(\mathcal{H}_0 (2\Omega-1)-10\big) (z+1)^{-2 (q+1)}}{36
\mathcal{H}_0^3}-\mathcal{H}_0 \big(\mathcal{H}_0 (2\Omega-7)-2\big) (z+1)^{2 q+2}
\\\label{63}
&-&18 \big(\mathcal{H}_0-1\big) \mathcal{H}_0 \big(\big
(\mathcal{H}_0-1\big) \mathcal{H}_0-4\big) \mu _2 (z+1)^{4
q+4}.
\end{eqnarray}
The plots of Figures \textbf{12} and \textbf{13} demonstrate the
behavior of energy density and pressure as functions of $z$ for all
considered models. The data indicates that the energy density
exhibits an upward trend and remains positive across different
values of $\Omega$ (Figure \textbf{12}). The pressure displays a
negative, decreasing trend and consistent with the behavior of
$\mathcal{DE}$ (Figure \textbf{13}).
\begin{figure}
\epsfig{file=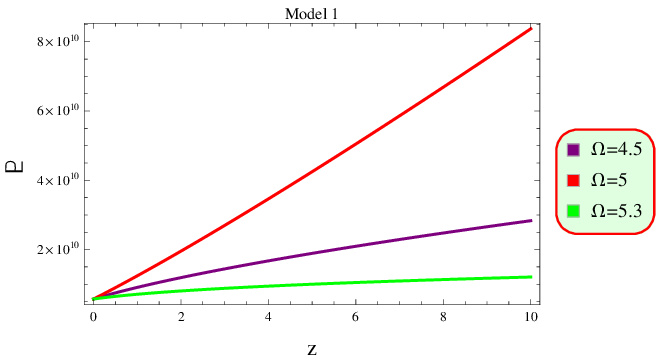,width=.5\linewidth}
\epsfig{file=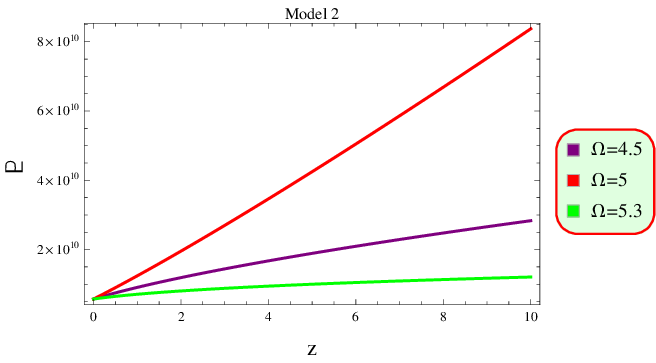,width=.5\linewidth}\center
\epsfig{file=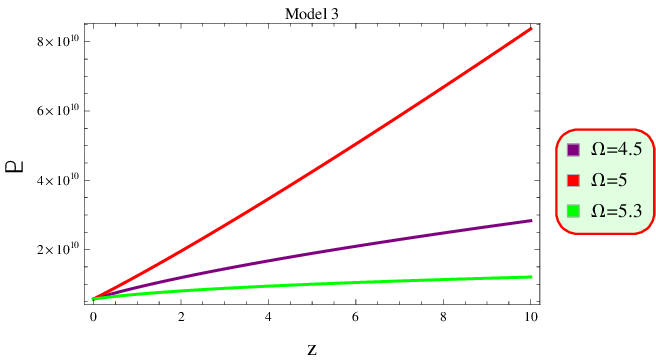,width=.5\linewidth}\caption{Behavior of energy
density versus redshift function for model \textbf{1}-\textbf{3}.}
\end{figure}
\begin{figure}
\epsfig{file=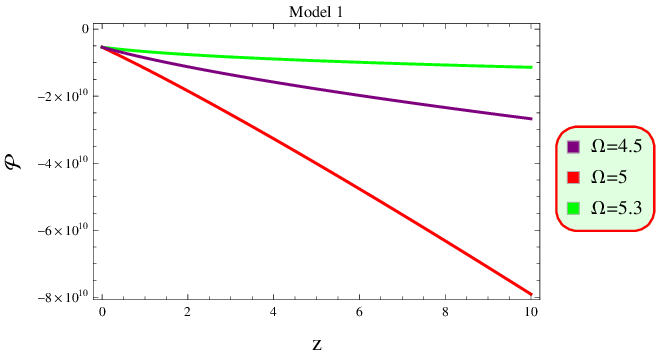,width=.5\linewidth}
\epsfig{file=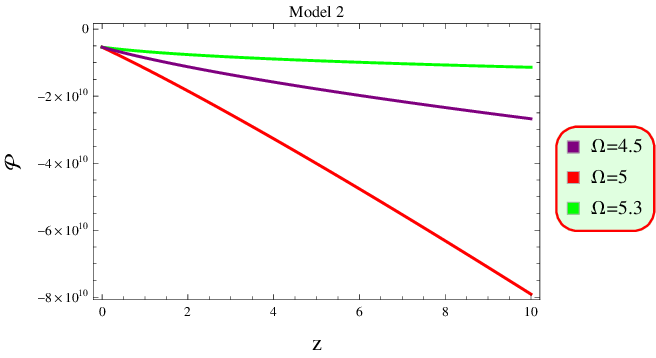,width=.5\linewidth}\center
\epsfig{file=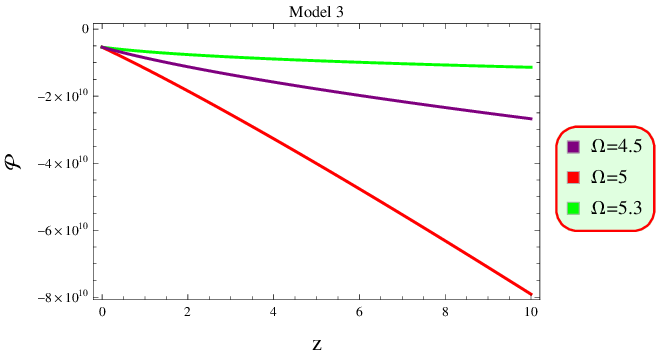,width=.5\linewidth}\caption{Behavior of
pressure with respect to redshift for different values of $\Omega$.}
\end{figure}

\section{Stability Analysis}

Understanding the stability of a cosmological model is essential for
analyzing the behavior of perturbations and predicting the
universe's evolution. This study conducts an analysis of the
stability of the Hubble parameter by assessing its response to
disturbances. To do this, we consider a linear perturbation of the
Hubble parameter as
\begin{equation}\label{44}
\mathcal{H}_{pert}(t)=\mathcal{H}(t)(\delta(t)_{\Omega}+1),
\end{equation}
where $\mathcal{H}(t)$ is the unperturbed Hubble parameter and
$\delta(t)_{\Omega}$ be a small perturbation. To facilitate our
analysis, we focus solely on perturbations of linear order.
Consequently, we assume $\delta(t)_{\Omega}$ as being smaller than
$\mathcal{H}(t)$, disregarding higher order terms. The conservation
equation for an effective fluid, which integrates the
characteristics of an ideal fluid, is essential for comprehending
the development of the cosmos in bouncing cosmology. This fluid has
several components, including matter, radiation, and $\mathcal{DE}$.
For an efficient fluid, the conservation equation may be articulated
as follows
\begin{equation}\label{45}
\dot{\varrho}+3H(t)(\varrho+\mathcal{P})=0.
\end{equation}
Utilizing Eqs. \eqref{29}, \eqref{30}, and \eqref{32}-\eqref{40} in
\eqref{45}, we derive the value of the perturbed term for models
\textbf{1}-\textbf{3} as follows
\begin{eqnarray}\nonumber
\delta_{\Omega}&=&-(2 (\Omega+2) \big(t^n \big(\vartheta _2+
\vartheta _3 t\big) +\vartheta _1 t\big) \big(-4
(\Omega-4) (\Omega+2)^2 \mu _2 t^{3 n} \big(\vartheta _2 +\vartheta _3 t\big){}^2
\\\nonumber
&\times&\big((n-1) \vartheta _2+n \vartheta _3 t\big)+2
(\Omega+2) t^{2 n} \big(\mu _2 \big(n \vartheta _3^2 t^2
\big(3 (3 \Omega+4) n-4 (\Omega-4)
\\\nonumber
&\times& (\Omega+2) \vartheta _1 t-3 (\Omega+4)\big)+(n-1)
\vartheta _2^2 \big(3 (3 \Omega n-4 \Omega+4n-8)-4
(\Omega
\\\nonumber
&-&4) (\Omega+2) \vartheta _1 t\big) +2 \vartheta _3
\vartheta _2 t \big(3 (n-1) (3 \Omega n-\Omega+4 n-4) -2
(\Omega-4) (\Omega
\\\nonumber
&+&2) (2 n-1) \vartheta _1 t\big)\big)+(3 \Omega-1) \mu _1 t^2
\big(\vartheta _2+ \vartheta _3 t\big){}^2\big)
+t^n \big(\vartheta _2 \big(\mu _1 t^2 \big((\Omega (4 \Omega
\\\nonumber
&+&13)+16) (n-1)+4 (\Omega+2) (3\Omega-1) \vartheta _1 t\big)
+2 (n-1) \mu _2 \big((\Omega+2) \vartheta _1 t
\\\nonumber
&\times& \big(3 (\Omega+4) (n-2)-2 (\Omega-4) (\Omega+2)
\vartheta _1 t\big)+9 \Omega (n-3)
(n-2)\big)\big)
\\\nonumber
&+&\vartheta _3 t \big(\mu _1 t^2 \big((\Omega (4 \Omega+13)+16)
n +4 (\Omega+2) (3 \Omega-1) \vartheta _1 t\big)
+2 n \mu _2
\\\nonumber
&\times& \big((\Omega+2) \vartheta _1 t \big(3 (\Omega+4)(n-1)-2
(\Omega-4) (\Omega+2) \vartheta _1 t\big)+9
\Omega (n-2)
\\\label{46}
&\times& (n-1)\big)\big)\big)+2 (\Omega+2) (3 \Omega-1)
\vartheta _1^2 \mu _1 t^4\big))(9 t^5)^{-1}.
\end{eqnarray}
For model \textbf{2}, the value of $\delta_{\Omega}$ is given as
\begin{eqnarray}\nonumber
\delta_{\Omega}&=&(2\ 3^{-\lambda -2} (\Omega+2)^2
\big(\big(\vartheta _2 +t \vartheta _3\big) \big(-4
3^{\lambda} (\Omega+2) \big(\vartheta _2+t \vartheta _3\big) \big(2 (\Omega+2)
\\\nonumber
&\times& \big(\vartheta _2+ t \vartheta _3\big) \big((n-1)
\vartheta _2+n t \vartheta _3\big) t^n+3 (n-1)
\big((n-2) \vartheta _2+n t \vartheta _3\big)\big)
\\\nonumber
&\times& \mu _2 t^n-(\lambda +1) \cosh (\lambda  \log (2))
\big(\frac{(\Omega+2)^2 \big(\big(\vartheta _2+ t
\vartheta _3\big) t^n+\vartheta _1 t\big){}^2}{t^2}\big){}^{\lambda }
\\\nonumber
&\times&\big((\Omega+2) (3\Omega-1) \big(\vartheta _2+t \vartheta _3\big){}^2 t^n+ 4 (\Omega+(\Omega-1)
\lambda+2) \big((n-1) \vartheta _2
\\\nonumber
&+&n t \vartheta _3\big)\big)\mu _3 t^2-(\lambda +1) \sinh (\lambda \log (2)) \big(\frac{(\Omega+2)^2
\big(\big(\vartheta _2+t \vartheta _3\big) t^n+\vartheta _1 t\big){}^2}{t^2}\big){}^{\lambda }
\\\nonumber
&\times&\big((\Omega+2) (3\Omega-1) \big(\vartheta _2+ t \vartheta
_3\big){}^2 t^n+4 (\Omega+(\Omega-1) \lambda +2) \big((n-1)
\vartheta _2
\\\nonumber
&+&n t \vartheta _3\big)\big) \mu _3 t^2\big) t^{2 n} +\vartheta _1
\big(3 (\Omega+2) \vartheta _2^2 \big(8\ 3^{\lambda }
\big(-(\Omega+2) (3 n-2) \vartheta _3 t^{n+1}
\\\nonumber
&-&n^2+3 n-2\big) \mu _2-2^{\lambda } (3 \Omega-1) t^2 (\lambda +1)
\big(\frac{(\Omega+2)^2 \big(\big(\vartheta _2+t \vartheta _3\big)
t^n+\vartheta _1 t\big){}^2}{t^2}\big){}^{\lambda }
\\\nonumber
&\times&\mu _3\big) t^n-8\ 3^{\lambda +1} (\Omega+2)^2 (n-1)
\vartheta _2^3 \mu _2 t^{2 n}+\vartheta _3 \big(-8 3^{\lambda +1}
(\Omega+2) n \vartheta _3
\\\nonumber
&\times&\big((\Omega+2) \vartheta _3 t^{n+1}+n-1\big) \mu _2
t^n-2^{\lambda } (\lambda +1) \big(3 (\Omega+2) (3 \Omega-1)
\vartheta _3 t^{n+1}
\\\nonumber
&+&4 n (\Omega+(\Omega-1) \lambda +2)\big) \big (\frac{(\Omega+2)^2
\big(\big (\vartheta _2+t \vartheta _3\big) t^n+\vartheta _1
t\big){}^2}{t^2}\big){}^{\lambda } \mu _3 t\big) t^2
\\\nonumber
&+&2 \vartheta _2 \big(4\ 3^{\lambda +1} (\Omega+2) t^n \vartheta _3
\big(-(\Omega+2) (3 n-1) \vartheta _3 t^{n+1}-2 (n-1)^2\big) \mu _2
\\\nonumber
&-&2^{\lambda } t (\lambda +1) \big(3 (\Omega+2) (3 \Omega-1)
\vartheta _3 t^{n+1}+ 2 (n-1) (\Omega+(\Omega-1) \lambda +2)\big)
\\\nonumber
&\times&\big(\frac{(\Omega+2)^2 \big(\big(\vartheta _2+ t \vartheta
_3\big) t^n+ \vartheta _1 t\big){}^2}{t^2}\big){}^{\lambda } \mu
_3\big) t\big) t^{n+1}-3 (\Omega+2) \vartheta _1^2 \big(2^{\lambda }
(3 \Omega-1)
\\\nonumber
&\times&t^2 (\lambda +1) \big(\vartheta _2+t \vartheta _3\big) \mu
_3 \big(\frac{(\Omega+2)^2 \big(\big(\vartheta _2+t \vartheta
_3\big) t^n+\vartheta _1 t\big){}^2}{t^2}\big){}^{\lambda }+4\
3^{\lambda }
\\\nonumber
&\times&\big(2 (\Omega+2) \big(\vartheta _2+ t \vartheta _3\big)
\big((n-1) \vartheta _2 +n t \vartheta _3\big) t^n+(n-1) \big((n-2)
\vartheta _2
\\\nonumber
&+&n t \vartheta _3\big)\big) \mu _2\big) t^{n+2}+ (\Omega+2)
\vartheta _1^3 \big(-8 3^{\lambda } (\Omega+2) \big((n-1) \vartheta
_2+n t \vartheta _3\big) \mu _2 t^n
\\\nonumber
&-&2^{\lambda } (3 \Omega-1) (\lambda +1) \big (\frac{(\Omega+2)^2
\big (\big(\vartheta _2+t \vartheta _3\big)
t^n+\vartheta _1 t\big){}^2}{t^2}\big){}^{\lambda } \mu _3 t^2\big) t^3\big))
\\\label{47}
&\times&(t^5)^{-1}.
\end{eqnarray}
For model \textbf{3}, $\delta_{\Omega}$ is defined as follows
\begin{eqnarray}\nonumber
\delta_{\Omega}&=&-(\big(\big(\vartheta _2+ t \vartheta _3\big)
t^n+ \vartheta _1 t\big) \big(4 (\Omega+2)^5
\big(\vartheta _2+t \vartheta _3\big){}^5 \big((3 \pi\Omega-1) \big(\vartheta _2
\\\nonumber
&+&t \vartheta _3\big) t^2 +12 \big((n-1) \big(n+4 (\Omega+2)
t \vartheta _1-2\big) \vartheta _2+ n t \big(n+4
(\Omega+2)
\\\nonumber
&\times&t\vartheta _1-1\big) \vartheta _3\big) \mu _2\big) t^{6
n}+32 (\Omega+2)^6 \big(\vartheta _2+t \vartheta _3\big){}^6
\big((n-1) \vartheta _2+n t \vartheta _3\big) \mu _2 t^{7 n}
\\\nonumber
&+&2 \big(16 (\Omega+2)^6 t \big((n-1) \vartheta _2 +n t
\vartheta _3\big) \mu _2 \vartheta _1^6+12 (\Omega+2)^5
\big((3 \Omega-1) \big(\vartheta _2+
\\\nonumber
&\times& t \vartheta _3\big) t^2+2 (n-1) \big((n-2) \vartheta _2 +n
t \vartheta _3\big) \mu _2\big) \vartheta _1^5+8 (\Omega+2)^5 t
\big((n-1) \vartheta _2
\\\nonumber
&+&n t \vartheta _3\big) \vartheta _1^4-9 (\Omega+2) (3 \Omega-1)
t^2\big(\vartheta _2+t \vartheta _3\big) \mu _4
\vartheta _1+18 (\Omega-4) t \big((n-1)
\\\nonumber
&\times& \vartheta _2+n t \vartheta _3\big) \mu _4\big) t^{n+5}
+(\Omega+2)\big(\vartheta _2+t \vartheta _3\big)
\big(192 (\Omega+2)^5 t \big((n-1) \vartheta _2
\\\nonumber
&+&n t \vartheta _3\big) \mu _2 \vartheta _1^5+60 (\Omega+2)^4
\big((3 \Omega-1) \big(\vartheta _2+t \vartheta
_3\big) t^2+4 (n-1) \big((n-2) \vartheta _2
\\\nonumber
&+&n t \vartheta _3\big) \mu _2\big) \vartheta _1^4
+64 (\Omega+2)^4 t \big((n-1) \vartheta _2+n t \vartheta
_3\big) \vartheta _1^3-9 (3 \Omega-1) t^2 \big(\vartheta _2
\\\nonumber
&+&t \vartheta _3\big) \mu _4\big) t^{2 n+4}+16 (\Omega+2)^5
\vartheta _1^2 \big(\vartheta _2+t \vartheta
_3\big){}^2 \big(\vartheta _2 \big(6 (n-1) t+5 \vartheta _1 \big((3\Omega
\\\nonumber
&-&1) t^2+6 (n-1) \big(n+(\Omega+2) t \vartheta _1-2\big)
\mu _2\big) \big)+t \vartheta _3 \big(6 n t+5 \vartheta
_1 \big((3 \Omega-1)
\\\nonumber
&\times&t^2+6 n \big(n+(\Omega+2) t \vartheta _1-1\big)
\mu _2\big)\big) \big) t^{3 n+3}+4 (\Omega+2)^5 \vartheta
_1 \big(\vartheta _2+t \vartheta _3\big){}^3
\\\nonumber
&\times&\big(\vartheta _2 \big(16 (n-1) t+5 \vartheta _1
\big(3 (3 \Omega-1) t^2+8 (n-1) \big(3 (n-2)+4
(\Omega+2)
\\\nonumber
&\times&t \vartheta _1\big) \mu _2\big)\big) +t \vartheta _3
\big(16 n t+5 \vartheta _1 \big(3 (3 \Omega-1)
t^2+8n\big(3 (n-1)+4 (\Omega+2)
\\\nonumber
&\times&t \vartheta _1\big) \mu _2\big)\big)\big) t^{4 n+2}
+8 (\Omega+2)^5 \big(\vartheta _2+t \vartheta
_3\big){}^4\big(\vartheta _2 \big(2 (n-1) t+3 \vartheta _1 \big((3 \Omega-
\\\nonumber
&\times& 1) t^2+10 (n-1) \big(n+2 (\Omega+2) t \vartheta _1
-2\big) \mu _2\big)\big)+t \vartheta _3 \big(2 n t+3
\vartheta _1 \big((3 \Omega
\\\nonumber
&-&1) t^2+10 n \big(n+2 (\Omega+2) t \vartheta _1-1\big)
\mu _2\big)\big)\big) t^{5 n+1}+(\Omega+2) (3 \Omega-1)
\vartheta _1^2
\\\nonumber
&\times&\big(4 (\Omega+2)^4 \vartheta _1^4-9 \mu _4\big)
t^8\big)) (18 (\Omega+2)^2 t^9 \big(\big(\vartheta _2+ t
\vartheta _3\big) t^{n-1}
\\\label{48}
&+&\vartheta _1\big){}^4)^{-1}.
\end{eqnarray}
\begin{figure}
\epsfig{file=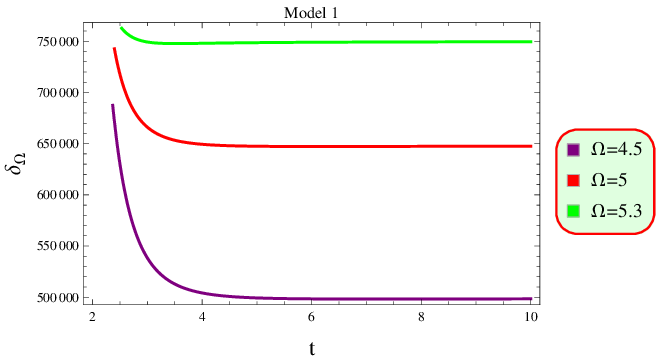,width=.5\linewidth}
\epsfig{file=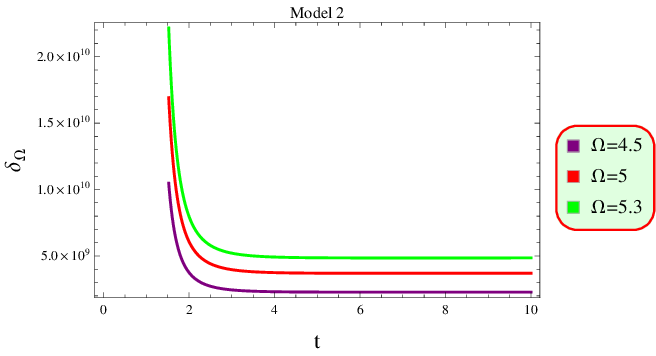,width=.5\linewidth}\center
\epsfig{file=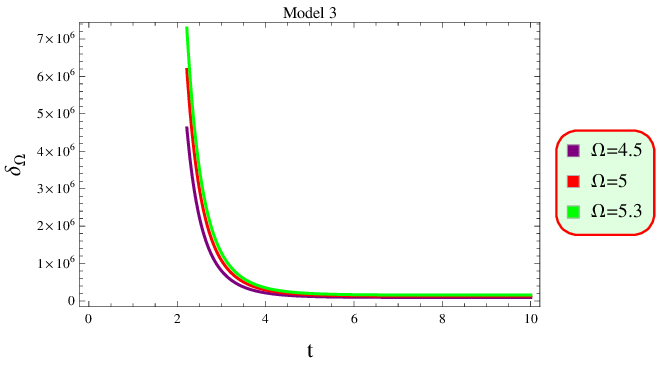,width=.5\linewidth}\caption{Behavior of
$\delta_{\Omega}$ over cosmic time for various values of $\Omega$.}
\end{figure}

The evolution of the perturbation term for all the models under
consideration throughout cosmic time for different values of
$\Omega$ is given in Figure \textbf{14}. At first, the perturbation
term shows variations but, as time progresses, these fluctuations
slowly decrease and approach to zero. This trend suggests that the
perturbation significance diminishes over time. As the value of
$\delta_{\Omega}$ decreases, the system achieves greater stability.
Ultimately, the stability of the cosmic structure in the later
stages relies on this particular behavior. By diminishing
perturbations, the cosmos remains stable and large scale
instabilities are prevented from growing and disrupting the
evolution of the universe.

\section{Final Remarks}

In recent years, limited data on the universe origin and evolution
have posed significant challenges for the scientific community. In
this perspective, cosmologists explored bouncing cosmology as an
alternative framework to address the limitations of the big bang
model, especially concerning issues like inflation and singularities
\cite{47}. This model provides a promising approach to address the
uncertainties around early universe singularities. The aim of this
study is to identify the unique features of a non singular bounce.
Here, we introduce an extension of $f(\mathcal{Q,C})$ gravity by
integrating the boundary term with non-metricity inside the
Lagrangian.

Bouncing cosmology investigates the concept that the cosmos goes
through cycles of expansion and contraction with a bounce occurring
when the universe transitions from a contracting phase to an
expanding one. One of the $\mathcal{MGT}s$ that has gained attention
in this context is the $f(\mathcal{Q}, \mathcal{C})$ theory.
Incorporating bouncing cosmology into the $f(\mathcal{Q},
\mathcal{C})$ theory introduces several novelties and motivations.
Bouncing cosmology offers a viable solution to the singularity
problem in standard cosmological models such as the Big Bang
singularity. By positing a bounce instead of a singularity, this
theory aims to provide a more complete and consistent description of
the universe evolution. The $f(\mathcal{Q}, \mathcal{C})$ theory
allows for the incorporation of $\mathcal{DE}$ and dark matter
effects in a unified framework. By modifying the gravitational
action through $f(\mathcal{Q}, \mathcal{C})$ theory can potentially
explain the accelerated expansion of the universe attributed to dark
energy as well as the gravitational effects attributed to dark
matter. Bouncing cosmology and $\mathcal{MGT}s$ often emerge from
attempts to reconcile $\mathcal{EGT}$ with quantum mechanics. By
exploring the consequences of $f(\mathcal{Q}, \mathcal{C})$ theory
in the context of bouncing cosmology, researchers aim to understand
how quantum gravitational effects might manifest at cosmological
scales, particularly near the bounce point.

In this study, we have identified several features of
$f(\mathcal{Q}, \mathcal{C})$ gravity that suggest it has the
potential to serve as a realistic alternative to $\mathcal{EGT}$.
One of the significant advantages of $f(\mathcal{Q}, \mathcal{C})$
gravity is its ability to provide non singular cosmological
solutions, particularly in the context of anisotropic models. The
existence of anisotropic bouncing solutions in the framework of
$f(\mathcal{Q}, \mathcal{C})$ gravity points to a mechanism that can
potentially resolve the big bang singularity problem. This feature
is crucial for any realistic cosmological model as it provides a
framework for smooth transitions between different phases of the
universe evolution without encountering singularities. While the
primary focus of our work has been on anisotropic solutions, the
broader class of cosmological solutions in $f(\mathcal{Q},
\mathcal{C})$ gravity, including those that mimic standard cosmology
at late times, suggests that this theory has the potential to be
compatible with current observational data.

It is important to clarify why $f(\mathcal{Q}, \mathcal{C})$ gravity
is specifically considered in this work as an alternative to
$\mathcal{EGT}$. While $\mathcal{EGT}$ remains the most successful
theory of gravity, there are several theoretical and observational
motivations for exploring alternatives like $f(\mathcal{Q},
\mathcal{C})$ gravity. The $f(\mathcal{Q}, \mathcal{C})$ gravity is
based on the symmetric teleparallel framework, where gravitational
interactions are governed by the non-metricity tensor, rather than
curvature (as in $\mathcal{EGT}$) or torsion (as in teleparallel
gravity). This provides a different geometric perspective on gravity
that may offer solutions to some of the unresolved problems in
$\mathcal{EGT}$, such as singularities and the need for
$\mathcal{DE}$ to explain cosmic acceleration. While the present
model primarily focuses on anisotropic bounce solutions and avoid
singularities, could have important implications for early universe
physics, including baryogenesis. In particular, the modifications to
the gravitational action in $f(\mathcal{Q}, \mathcal{C})$ gravity
may introduce new mechanisms for generating the observed matter,
antimatter asymmetry, though this requires further investigation.
This suggests that $f(\mathcal{Q}, \mathcal{C})$ gravity is not only
a mathematically interesting alternative but also a potentially
viable one in cosmological contexts. In the revised manuscript, we
have expand on the physical motivations for considering
$f(\mathcal{Q}, \mathcal{C})$ gravity, emphasizing its unique
features and the potential it offers for resolving certain issues in
$\mathcal{EGT}$, such as singularities and anisotropic evolution.

The key findings of our investigation are summarized as follows:
\begin{itemize}
\item
The scale factor shows positively increasing behavior which
represents that the universe is exhibiting increasing expansion
scenario (Figure \textbf{1}). Additionally, the parametric values
$\vartheta_{1}=-0.05$ and $n=2.3$ are selected to effectively
capture the cosmological bounce solutions. These parameters
guarantee that the $f(\mathcal{Q,C})$ model corresponds with both
theoretical predictions and empirical data, precisely representing
the dynamics of cosmic expansion.
\item
The Hubble parameter provides an effective way to characterize
different phases in the bounce model, where $\mathcal{H}<0$
corresponds to the contraction phase, $\mathcal{H}=0$ signifies the
transition or bounce point and $\mathcal{H}>0$ indicates the
expansion phase. This demonstrates the model smooth dynamical
behavior. By analyzing the time derivative of the Hubble parameter
at $t=8.912$, one can observe the universe shift from contraction to
expansion (Figure \textbf{2}).
\item
The parameters $\vartheta_2$ and $\vartheta_3$ are utilized
to generate the oscillatory behavior observed between
the contraction phase $(\mathcal{H} < 0)$ and the
expansion phase $(\mathcal{H} > 0)$ in the current universe
(refer to Tables \textbf{1} and \textbf{2}).
\item
The cosmos is seen to grow at an accelerating rate,
reaching its maximum value at $t = 8.912$ (Figure
\textbf{3}).
\item
The energy density exhibits a positive upward trend, whereas the
pressure stays negative throughout all models, consistent with the
behavior of the $\mathcal{DE}$ model (Figures
\textbf{4}-\textbf{6}).
\item
The $\mathcal{EOS}$ represents the phantom regime, aligning with
predictions of accelerated cosmic dynamics (Figure \textbf{7}).
\item
The $\mathcal{NEC}$ exhibits negative behavior for all considered
models which indicates the presence of non singular bounce in
$f(\mathcal{Q,C})$ theory (Figures \textbf{8}-\textbf{10}).
\item
The Hubble radius demonstrates a gradual shift from contraction to
expansion phases (Figure \textbf{11}). This transition is shaped by
the cosmic timeline leading up to the bounce and experiences changes
as time progresses beyond the bounce point.
\item
The graphical representation of matter configuration and
cosmological parameters versus redshift function show rapid
expansion of the cosmos.(Figure \textbf{12} and \textbf{13})
\item
Stability analysis demonstrates that the perturbation parameters
remain stable over time, indicating the model resilience to small
fluctuations and affirming its reliability as a framework for
understanding cosmic evolution (Figure \textbf{14}).
\end{itemize}

This research has examined bouncing cosmology in $f(\mathcal{Q},
\mathcal{C})$ gravity, which provides important perspectives for
future investigations into the origins of the universe. One of the
most notable observations is the meticulous examination of viable
bouncing solutions. This approach has the potential to address
challenges such as the singularity problem and offer an alternative
viewpoint on the phenomenon of early cosmic dynamics. The paper
provides an in depth analysis and imposes constraints on various
cosmological parameters in the $f(\mathcal{Q}, \mathcal{C})$
framework. Agrawal et al \cite{39} studied an extended gravitational
theory that could create viable models of the universe to address
the issue of late time expansion. The variation of energy content
and pressure in relation to the redshift function has been studied
in several $\mathcal{MGT}s$ \cite{48}. Sharif et al \cite{50}
examined this correlation for $f(\mathcal{Q})$ models utilizing
Bianchi type-1 spacetime. We have observed more favorable evaluation
of the matter variables in relation to the redshift function in
$f(\mathcal{Q}, \mathcal{C})$ gravity.\\\\
\textbf{Data Availability Statement:} No new data were produced or
examined in relation to this study.

\end{document}